\documentclass[fleqn,usenatbib]{mnras}
\usepackage{newtxtext,newtxmath}
\usepackage{ae,aecompl}
\usepackage[utf8]{inputenc}
\usepackage{graphicx}	
\usepackage{amsmath}	
\usepackage{amssymb}
\usepackage{xcolor}
\newcommand{\iu}{{i\mkern1mu}}
\begin{document}
\title{Low-Eccentricity Migration of Ultra-Short Period Planets in Multi-Planet Systems}

\author[Pu \& Lai]{
Bonan Pu \thanks{E-mail: bonanpu@astro.cornell.edu (BP)} and Dong Lai
\\
Cornell Center for Astrophysics and Planetary Science, Department of Astronomy, 
Cornell University, Ithaca, NY 14853, USA\\
}
\maketitle
\begin{abstract} 
Recent studies suggest that ultra-short period planets (USPs),
Earth-sized planets with sub-day periods, constitute a statistically
distinct sub-sample of {\it Kepler} planets: USPs have smaller radii
($1-1.4R_\oplus$) and larger mutual inclinations with neighboring
planets than nominal {\it Kepler} planets, and their period
distribution is steeper than longer-period planets.  We study a
"low-eccentricity" migration scenario for the formation of USPs, in
which a low-mass planet with initial period of a few days maintains a
small but finite eccentricity due to secular forcings from exterior
companion planets, and experiences orbital decay due to tidal
dissipation.  USP formation in this scenario requires that the initial
multi-planet system have modest eccentricities ($\gtrsim 0.1$) or
angular momentum deficit. During the orbital decay of the inner-most
planet, the system can encounter several apsidal and nodal precession
resonances that significantly enhance eccentricity excitation and
increase the mutual inclination between the inner planets.  We develop
an approximate method based on eccentricity and inclination eigenmodes
to efficiently evolve a large number of multi-planet systems over Gyr
timescales in the presence of rapid (as short as $\sim 100$~years)
secular planet-planet interactions and other short-range forces.
Through a population synthesis calculation, we demonstrate that the
"low-$e$ migration" mechanism can naturally produce USPs from the
large population of {\it Kepler} multis under a variety of conditions,
with little fine tuning of parameters. This mechanism favors smaller
inner planets with more massive and eccentric companion planets, and
the resulting USPs have properties that are consistent with
observations.
\end{abstract}

\section{Introduction}
\label{sec:1}
The existence of ultra-short period planets (USPs), defined to be small planets $(R \le 2R_{\oplus}$), with sub-day periods (i.e. $P \le 1$ day) is one of the major surprises in exoplanetary astrophysics. The first example of such planets, CoRoT-7b, was discovered in 2009 \citep{Leger2009}; since then, about a hundred USPs have been found by various transit surveys \citep{Sanchis-Ojeda2014}, and the overall occurence rate of USPs sits at $\sim 1\%$, a figure that is similar to the census of hot Jupiters, i.e. giant planets with $P \le 10$ days \citep{Cumming2008,Wright2012}. A few notable USPs have received special attention: 55 Cnc e \citep{Dawson2010} with $P = 0.74$ days was the first discovered Super-Earth, Kepler-10b with $P = 0.83$ days \citep{Batalha2011} was the first terrestrial planet discovered by {\it Kepler}, and Kepler-78b \citep{Sanchis2013} with $P = 0.36$ days is the current record holder amongst planets known with the shortest orbital period. Kepler-32 and Kepler-80 are another pair of unusually interesting systems: both contain a USP with an additional set of three or four exterior transiting planets that are potentially locked in mean-motion resonances \citep{Swift2013,MacDonald2016}. A recent review on the detection and population statistics of USPs is provided by \cite{Winn2018}. 

Historically, the choice of the one day cut-off for the definition of USPs was not astrophysically motivated; it was chosen because the number of planets discovered below such period was rare at the time \citep{Winn2018}. However, recent evidence has emerged that USPs may in fact be a statistically distinct population.
Planets with $P \le 1$ days appear to follow a different period distribution than planets above the one day cut-off: \cite{Lee2017} found that whereas transiting planets with $1 \le P \le 10$ days followed a power law $dN/d \log P \propto P^{\alpha}$ with $\alpha \simeq 1.5 - 1.7$ \citep[see also][]{Petigura2018, Weiss2018}, USPs followed a steeper trend with $\alpha \sim 3.0$. In addition, the normalization of the period distribution may also be different: the planet occurrence rate is discontinuous across the $P = 1$ day boundary, with $\sim$ 50\% more planets with periods just below $P = 1$ days than just above. 

In addition to their period distribution, USPs have other statistical properties that differentiate them from longer-period planets. USPs have smaller radii, with the vast majority having $1 R_{\oplus} \le R \le 1.4 R_{\oplus}$ \citep{Winn2018}, a fact which may be attributed to photo-evaporation or `boil-off' as the planets are intensely irradiated. 
Compared with the other {\it Kepler} planets, planet systems with USPs also appear to have higher mutual inclinations: \cite{Dai2018} found that transiting {\it Kepler} planets with a semi-major axis to stellar radius ratio $a/R_{\star} < 5$ had an inclination dispersion of $\Delta \theta \approx 6.7 \pm 0.7$ degrees, while planets with $5 < a/R_{\star} < 12$ had only $\Delta \theta \approx 2.0 \pm 0.1$ \cite[consistent with the overall figures for Kepler multis, see e.g. ][]{Tremaine2012, Fang2012, Fabrycky2014}. This observation is further corroborated by the fact that for FGK host stars, USPs feature a factor of $\sim 8$ fewer co-transiting external companions compared with their merely `short-period planet' (SP) counterparts \citep{Petrovich2018, Weiss2018}, and when USPs do have external transiting companions, the period ratios between the USP and their closest companion is $P_2/P_1 \gtrsim 15$, a value that is nearly an order of magnitude above the typical period ratios of $1.3 - 4.0$ seen in {\it Kepler} multis \citep[see also][]{Steffen2013}.

The statistical distinctness of USPs and their unusual locations so close to their host stars defy conventional understandings of planet formation, and the origins of these planets remains a mystery. USPs may sit in the short-period tail of the distribution of close-in rocky planets that formed in-situ through core accretion \citep{Chiang2013}, or they may have migrated to their current locations from initially more distant orbits \citep{Ida2004,Schlaufman2010,Terquem2014}. In the latter scenario, they (like most of their SP bretheren) likely would have begun their lives as Super-Earths/Mini-Neptunes with a gaseous H/He envelope that was subsequently lost to photo-evaporation \citep{Valencia2010,Owen2013}. To shove the planets very close to their host stars, some form of disk migration and/or tidal dissipation is required. \cite{Lee2017} considered stellar tides raised by the planet, treating the stellar tidal quality factor $Q'_{\star}$ as a free parameter; since the orbital decay rate is proportional to the planet mass, the required $Q'_{\star}$ value to induce significant decay of small planets would make hot Jupiters at $P \sim 1$ day ``disappear'' on a short timescale. In the case of planetary tides, the proto-USP must maintain a finite eccentricity in order to undergo orbital decay. \cite{Petrovich2018} examined a high-eccentricity migration scenario in which a proto-USP attains large eccentricity due to secular chaos in a hierarchical system with $N > 3$ planets, followed by orbital circularization due to planetary tides. They also briefly explored the possibility of forming USPs through secular interactions with eccentric giant planet companions, but dismissed the possibility as unlikely: they found that producing USPs usually required eccentric giant planet companions with $P \le 10$ days, a requirement at odds with the observation that the presence of USPs do not seem to be correlated with the stellar host metallicity, and therefore by proxy the occurrence of giant planet companions \citep{Winn2017}. In short, although these previous ideas indeed may produce ultra-short period planets under some conditions or assumptions, there is yet no firm evidence that USP formation can be completely accounted for by any one of the aforementioned scenarios. 

Indeed, the formation mechanism of USPs remains unclear and this is the question we aim to address. The main thesis of this work is that the combination of secular interactions and tidal dissipation in multi-planet systems is likely be the most natural and efficient way to generate USPs. This mechanism requires small, rocky planets to be born at moderate eccentricities (i.e. $e \gtrsim 0.1$) in multi-planet ($N \ge 3$) systems, but otherwise requires little fine tuning of planet parameters. Empirical studies suggest the orbital eccentricities of Kepler multis have a dispersion of $\sigma_e \sim 0.05$ \citep{Xie2016,vanEylen2015,VanEylen2018}, so USP formation in this mechanism would occur at the tail end of the eccentricity distribution. However, note that the currently observed eccentricity distribution has been damped over Gyrs by tidal dissipation \citep{Hansen2015a}, and the primordial eccentricities may be much larger. 

Certainly, the idea of secular forcings coupled with tidal dissipation is not a new one; it has already been applied to short-period exoplanet systems in various contexts  \citep[e.g.][]{Wu2002, Mardling2007, Mardling2010, Batygin2009, Hansen2015b, Petrovich2018}, although this work is the first to tackle the problem in the context of USP formation in multi-planet systems. The mechanism studied in this paper has some similarity to the secular chaos mechanism proposed by \cite{Petrovich2018}, but with important differences: Whereas \cite{Petrovich2018} rely on secular chaos driving small planets to attain large eccentricities (e.g. $1-e \ll 1$) and thereby small pericenter distances to achieve USP formation, our mechanism requires the inner planet (initially at $P = 1 - 3$ days) to achieve only mild initial eccentricities ($e \sim 0.1$) through secular interactions; although the mechanism proposed by \cite{Petrovich2018} allows for a more diverse proto-USP period, it also requires the presence of several well-separated exterior planets, whereas in our mechanism, the initial proto-USP period is more constrained, but the external planet companions are allowed more lee-way in terms of their spacing. In light of this fact, we call our proposed formation mechanism the \color{black}`low-$e$ migration' of USPs.\color{black}

In this paper, we present an investigation on the prospects of generating USPs through \color{black}`low-$e$ migration'\color{black}. The structure of the paper is as follows. In section \ref{sec:2}, we present the mathematical formalism for the eccentricity and orbital evolution of a multi-planet system undergoing secular interactions and tidal dissipation. As we discuss below, a brute-force approach to this problem is impractical, and we derive the evolution equations in the framework of eigenmodes in section \ref{sec:2.1}. In section \ref{sec:2.2}, we apply our formalism to the case of 2-planet systems, deriving semi-analytical results for the eccentricity and mode evolution; in section \ref{sec:2.3} we discuss the criterion for USP formation to occur in such 2-planet systems. We then extend these results to the case of 3-planet systems in section \ref{sec:3}, and show that such systems allow for successful low-$e$ secular migration under reasonable conditions. In section \ref{sec:4} we consider the inclination evolution of the planets, taking into account planet-planet coupling as well as interactions with stellar spin. The results of the preceding sections are synthesized into a population model in section \ref{sec:5} - readers who are most interested in observational implications of our results may skip to this section. In section \ref{sec:6}, we discuss the feasibility of low-$e$ secular migration and compare it against other proposed mechanisms. Finally, a summary of our work is provided in section \ref{sec:7}. 

\section{Eccentricity Evolution and Orbital Decay: Formalism}
\label{sec:2}
Consider a N-planet system with individual planet mass $m_i$, semi-major axis $a_i$, initial orbital eccentricity $e_i$, longitude of periapsis $\varpi_{i}$, inclination $\theta_i$ and longitude of the ascending node $\Omega_i$, where $i \in [1, ~N]$ is the planet index, orbiting a host star with mass $M_{\star}$ and radius $R_{\star}$. We assume that the host-star is Sun-like, i.e. $M_{\star} = M_{\odot}$ and $R_{\star} = R_{\odot}$. The planet's semi-major axis is related to the orbital period by $a \simeq 0.0196 (P/\mathrm{day})^{2/3} (M_{\star}/M_{\odot})^{1/3}$ au. The dynamical evolution of the system is governed by the interplay of several effects: planet-planet secular perturbations, General Relativistic (GR) periastron advance, spin-orbit coupling due to stellar oblateness, planetary tides and stellar tides. 
We define $\mathcal{E}_i \equiv e_i \exp{(\iu \varpi_i)}$ to be the complex eccentricity of the $i$-th planet, and define the eccentricity vector of the N-planet system as
\begin{equation}
    \vec{\mathcal{E}} = \begin{pmatrix} \mathcal{E}_1 \\ \mathcal{E}_2 \\ \vdots  \end{pmatrix}.
\end{equation}
In the linear (Laplace-Lagrange) theory, the time evolution of $\vec{\mathcal{E}}$ is governed by the equation
\begin{align}
\frac{d}{dt} \vec{\mathcal{E}}(t) &= \iu \mathbf{H}(t) \vec{\mathcal{E}}(t),
  \label{eq:dEdt}
\end{align}
where the coefficients of the time-varying $N\times N$ matrix $\mathbf{H}(t)$ is given by
\begin{equation}
H(t) = 
 \begin{pmatrix}
  \tilde{\omega}_{1} & -\nu_{12} & \cdots & -\nu_{1N}  \\
  -\nu_{21} & \tilde{\omega}_{2} & \cdots & -\nu_{2N}   \\
  \vdots  & \vdots  & \ddots & \vdots \\
  -\nu_{N1} & -\nu_{N2} & \cdots & \tilde{\omega}_{N} \end{pmatrix} .
\label{eq:Ht}
\end{equation}
Here the complex ``frequencies'' $\tilde{\omega}_{i}$ (taking into account the eccentricity damping due to tidal dissipation in the $i$-th planet) is defined as
\begin{equation}
\tilde{\omega}_i \equiv \omega_i + \iu \gamma_i = \sum_{j \neq i} \omega_{ij} + \omega_{i,\mathrm{gr}} + \omega_{i,\mathrm{tide}} + \iu \gamma_i.
\end{equation}
The quantities $\omega_{ij}$ and $\nu_{ij}$ are the quadrupole and
octupole precession frequencies of the $i$-th planet driven by the actions of the $j$-th planet, given by
\begin{align}
\omega_{ij} &= \frac{G m_i m_j a_{<}}{4a_{>}^2 L_i } b^{(1)}_{3/2}(\alpha) \nonumber \\ &\simeq 
4.0 \times 10^{-4} 
\left(\frac{M_{\star}}{M_{\odot}}\right)^{-1/2}
\left(\frac{m_j}{10 M_{\oplus}}\right) 
\left(\frac{a_i}{0.02 ~\mathrm{au}}\right)^{3/2}  \nonumber  \\ &~~~~ \times
\left(\frac{a_j}{0.1 ~\mathrm{au}}\right)^{-3}
\left( \frac{b_{3/2}^{(1)}(\alpha)}{3\alpha} \right) ~\mathrm{yr}^{-1}  ,
\label{eq:wjk}
\end{align}
and
\begin{align}
\nu_{ij} &= \frac{G m_i m_j a_{<} }{4a_{>}^2 L_i} b^{(2)}_{3/2}(\alpha)  \nonumber  \\ &\simeq
1.0 \times 10^{-4} 
\left(\frac{M_{\star}}{M_{\odot}}\right)^{-1/2}
\left(\frac{m_j}{10 M_{\oplus}}\right) 
\left(\frac{a_i}{0.02 ~\mathrm{au}}\right)^{3/2}  \nonumber  \\ &~~~~ \times
\left(\frac{a_j}{0.1 ~\mathrm{au}}\right)^{-3}
\left( \frac{b_{3/2}^{(2)}(\alpha)}{15\alpha^2/4} \right) ~\mathrm{yr}^{-1} 
\label{eq:vjk},
\end{align}
where we have defined $a_< = \mathrm{min}(a_i,a_j)$, $a_> = \mathrm{max}(a_i,a_j)$, $\alpha = a_< / a_>$,
$L_i = m_i\sqrt{GM_*a_i}$ is the (circular) angular momentum of the $i$-th planet, and $b_{3/2}^{(n)}(\alpha)$ are the usual Laplace coefficients given by
\begin{equation}
    b^{(n)}_{3/2}(\alpha) = \frac{1}{2\pi} \int_{0}^{\pi} \frac{\cos{(nt)}}{(\alpha^2 + 1 - 2\alpha \cos{t})^{3/2}} dt.
    \label{eq:b32n}
\end{equation}
In the limit $\alpha \ll 1$, the first-order expansion of the Laplace coefficients are $b_{3/2}^{(2)}(\alpha) \simeq 3\alpha$ and $b_{3/2}^{(2)}(\alpha) \simeq 15 \alpha^2/4$.

The general relativistic apsidal precession frequency is given by (for $e_i \ll 1$):
\begin{equation}
\omega_{i,\mathrm{gr}} = \frac{3GM_{\star}}{c^2 a_i} n_i \simeq
3.3\times 10^{-3} 
\left(\frac{M_{\star}}{M_{\odot}}\right)^{3/2}
\left(\frac{a_i}{0.02 \mathrm{au}}\right)^{-5/2} ~\mathrm{yr}^{-1} , 
\end{equation}
where $n_i$ is the angular orbital frequency. The rate of periastron advance on the $i$-th planet due to its tidal bulge is given by
\begin{align}
\omega_{i,\mathrm{tide}} &=  \frac{15}{2}k_{2,i} \frac{M_{\star}}{m_i}\left(\frac{R_i}{a_i} \right)^{5} n_i \nonumber \\
&= 2.44 \times 10^{-4}  k_{2,i} 
\left(\frac{M_{\star}}{M_{\odot}}\right)^{1.5}
\left(\frac{m_i}{M_{\oplus}}\right)
\left(\frac{R_i}{R_{\oplus}}\right)^5 \nonumber \\ &~~~~\times
\left(\frac{a_i}{0.02 ~\mathrm{au}}\right)^{-13/2} ~\mathrm{yr}^{-1},
\end{align}
with $k_{2,i}$ being the tidal Love number of the $i$-th planet; in this work, we adopt a value of $k_{2,i} = 1$. Generally, as the planet moves inwards, the GR and tidal forces become increasingly important, while farther out, planet-planet secular interactions tend to prevail. 

We use the weak friction theory of equilibrium tides to describe tidal dissipation in the planet \citep{Darwin1880, Alexander1973, Hut1981}. The eccentricity damping rate of the $i$-th planet due to tidal dissipation is 
\begin{align}
\left(\frac{\dot{e}_i}{e_i} \right)_{\mathrm{tide}} &\equiv -\gamma_i = -\frac{21}{2}k_{2,i} \Delta t_{L,i} \frac{M_{\star}}{M_{\odot}} \left(\frac{R_i}{a_i} \right)^{5} n_i^2  \nonumber \\
&= -2.4\times 10^{-6}  \times  k_{2,i} \left(\frac{\Delta t_{L,i}}{100 \mathrm{s}}\right) 
\left(\frac{M_{\star}}{M_{\odot}}\right)^2 \nonumber \\
&~~\times \left(\frac{m_i}{M_{\oplus}}\right)^{-1}
\left(\frac{R_i}{R_{\oplus}}\right)^{5}
\left(\frac{a_i}{0.02 ~\mathrm{au}}\right)^{-8} ~\mathrm{yr}^{-1},
\end{align}
where $\Delta t_{L,i}$ is the tidal lag time of the $i$-th planet, and is related to the tidal quality factor $Q_i$ by
\begin{equation}
    Q_i = (2n_i\Delta t_{L,i})^{-1} = 70 \left(\frac{a_i}{0.02 \mathrm{au}} \right)^{3/2} \left(\frac{M_{\star}}{M_{\odot}} \right)^{-1/2} \left(\frac{\Delta t_{L,i}}{100 \mathrm{s}} \right)^{-1}.
\end{equation}
Note that in this formalism, the value of $Q_i$ is not constant and instead varies with the planet's semi-major axis. In the Solar System, values of $Q$ range from $10 - 500$ for terrestrial planets and satillites, but the gas giants have values of $Q$ that are much larger \citep{Goldreich1996, Ogilvie2014}. We assume proto-USPs to be predominantly rocky and adopt values of $Q_1$ in the range between 70 and 700, while the exterior planets are assumed to have H/He envelopes comprising a few percent of the planet's mass (but a large fraction of the radius) and therefore have much larger values of tidal $Q_i$.

Because the the outer planets ($i \ge 2$) have much larger values of $a_i$ and $Q_i$, the effects of tidal dissipation are much weaker for these planets. Therefore, we simplify the problem by only considering tidal dissipation for the innermost planet (i.e. by setting $Q_i = \infty$ for $i \ge 2$) in sections \ref{sec:2} through \ref{sec:4}; the effects of tidal dissipation in the outer planets are included in our population synthesis study in section \ref{sec:5}.

To completely determine the time evolution of the system, Eq. (\ref{eq:dEdt}) for $\vec{\mathcal{E}}$ should be supplemented by the evolution of the inner planet's semi-major axis $a_1$:
\begin{align}
\left(\frac{\dot{a}_1}{a_1} \right)_{\mathrm{tide}} &= -2\gamma_1 e_1^2
=  -1.9 \times 10^{-9} k_{2,1}
\left(\frac{\Delta t_{L,1}}{100 \mathrm{s}}\right)
\left(\frac{e_1}{0.02}\right)^{2} \nonumber \\ &~~ \times 
\left(\frac{M_{\star}}{M_{\odot}}\right)^{2}
\left(\frac{m_1}{M_{\oplus}}\right)^{-1}
\left(\frac{R_1}{R_{\oplus}}\right)^{5}
\left(\frac{a_1}{0.02 ~\mathrm{au}}\right)^{-8} ~ \mathrm{yr^{-1}} .
\label{eq:adot_tide}
\end{align}
We also consider the orbital decay driven by dissipation of tides raised on the host star by the planet \citep{Goldreich1996}:
\begin{align}
\left(\frac{\dot{a}_1}{a_1} \right)_{\mathrm{tide}\star} &\equiv -\gamma_{\star}  = -\frac{9}{2} \left(\frac{m_1}{M_{\star}} \right) \left(\frac{R_{\star}}{a_1}\right)^{5} \frac{n_1}{Q'_{\star}}  \nonumber \\
&=  -1.85 \times 10^{-9} 
\left(\frac{M_{\star}}{M_{\odot}}\right)^{-1/2}
\left(\frac{R_{\star}}{R_{\odot}}\right)^5 \nonumber 
\left(\frac{Q'_{\star}}{10^6}\right)^{-1} \\ &~~~~~~~~~ \times
\left(\frac{m_1}{M_{\oplus}}\right)
\left(\frac{a_1}{0.01 ~\mathrm{au}}\right)^{-13/2} ~\mathrm{yr}^{-1},
\label{eq:gamma_star}
\end{align}
where $Q'_{\star} = 3Q_{\star} / (2k_{2,\star})$ is the reduced tidal quality factor of the star. Empirical measurements by \cite{Penev2018} suggest a value of $Q'_{\star} = 10^7$ at a tidal forcing frequency of 2 days$^{-1}$, decreasing to $Q'_{\star} = 10^5$ when the forcing frequency is 0.5 day$^{-1}$. \cite{Lee2017} treated $Q'_{\star}$ as a free parameter, and considered values of $Q'_{\star}$ in the range of $10^{6} - 10^{8}$.
Thus, in general, the orbital decay rate of the inner-most planet is given by
\begin{equation}
\dot{a}_1 = -2 \gamma_{1} |\mathcal{E}_1|^2 a_1 - \gamma_{\star} a_1.
\label{eq:dadt}
\end{equation}
In sections \ref{sec:2} - \ref{sec:4} we will focus on planetary tides and neglect the effect of stellar tides (i.e. by setting $Q'_{\star} = \infty)$, although we will include stellar tides in our population synthesis study in section \ref{sec:5}. We do this for two reasons: Firstly,  for typical values of $Q'_{\star}$ the effect of stellar tides is small over the lifetime of the system and only attains significance when an USP has already been produced (i.e. $a_1 \le 0.02$ au.), therefore its role is orthogonal to the aims of this work. Secondly, the addition of stellar tidal dissipation destroys the conservation of orbital angular momentum, an otherwise desirable property of Eqs. (\ref{eq:dEdt}) \& (\ref{eq:dadt}), as we will demonstrate in section \ref{sec:2.3}.

\subsection{Eccentrity Evolution in the Framework of Eigenmodes}
\label{sec:2.1}
A brute-force integration of Eqs. (\ref{eq:dEdt}) and (\ref{eq:dadt}) encounters difficulty: the relevant frequencies vary over many orders of magnitude with the orbital decay timescale  ($|a_1/\dot{a}_1| \gtrsim 1$ Gyr) being much longer than the precession timescales ($2\pi/\omega_{i} \sim 10^3$ yrs); the ``stiffness'' of the equations make it impractical to integrate a large number of systems. Our approach is eschew calculating the phase of the eccentricity vector. We do this by decomposing the planet eccentricities into eigenmodes. 

We define the eigenvalue $\lambda_{\alpha}$ and eigenvector $\vec{\mathcal{E}}_{\alpha}$ (with modes denoted using Roman Numerals $\alpha \in [\mathrm{I, ~II, ~III...}]$) of the system as
\begin{equation}
    \mathbf{H}(t)\mathcal{E}_{\alpha}(t) = \lambda_{\alpha}(t) \vec{\mathcal{E}}_\alpha(t),
\end{equation}
where
\begin{equation}
    \vec{\mathcal{E}}_\alpha = \begin{pmatrix} \mathcal{E}_{\alpha1} \\ \mathcal{E}_{\alpha2} \\ \vdots \end{pmatrix}.
\end{equation}
Note that since $\mathbf{H}(t)$ depends on $a_1$ (the time evolution of $a_2, ~a_3...$ are negligible), the eigenvalue $\lambda_{\alpha}(t)$ and eigenvector $\vec{\mathcal{E}}_\alpha(t)$ evolve in time as $a_1$ decreases. We now introduce the matrices $\mathbf{G}(t)$ and $\mathbf{V}(t)$ formed from the eigenvalues and eigenvectors of $\mathbf{H}(t)$:
\begin{equation}
\mathbf{G}(t) = 
\mathrm{diag}(\lambda_{\mathrm{I}}, \lambda_{\mathrm{II}}, \hdots, \lambda_{\mathrm{N}})
  \label{eq:gmat}
\end{equation}
and
\begin{equation}
\mathbf{V}(t) = 
\begin{bmatrix}
 \vec{\mathcal{E}}_{\mathrm{I}}  & \vec{\mathcal{E}}_{\mathrm{II}}  & \hdots & \vec{\mathcal{E}}_{\mathrm{N}} 
\end{bmatrix}.
\end{equation}
By definition, the matrices $\mathbf{G}(t)$ and $\mathbf{V}(t)$ satisfy the identity
\begin{equation}
    \mathbf{G}(t) = \mathbf{V}^{-1}(t) \mathbf{H}(t) \mathbf{V}(t).
    \label{eq:GVHV}
\end{equation}
In general, the time evolution of $\vec{\mathcal{E}}$ can be written as a superposition of eigenmodes,
\begin{equation}
    \vec{\mathcal{E}}(t) = \sum_\alpha A_{\alpha}(t) \vec{\mathcal{E}}_{\alpha}(t) = \mathbf{V}(t) \vec{A}(t),
    \label{eq:et_decomp}
\end{equation}
where $\vec{A}(t) \in \mathbb{C}^N$ is the vector of eigenmode amplitudes:
\begin{equation}
    \vec{A}(t) \equiv \begin{pmatrix} A_{\mathrm{I}} \\ A_{\mathrm{II}} \\ \vdots \end{pmatrix}.
\end{equation}
The initial condition $\vec{A}(0)$ is given by
\begin{equation}
    \vec{A}(0) = \mathbf{V}^{-1}(0) ~\vec{\mathcal{E}}(0).
\end{equation}
Substituting Eq. (\ref{eq:et_decomp}) into Eq. (\ref{eq:dEdt}), and using the identity $\mathbf{H}\mathbf{V}\vec{A} = \mathbf{V}\mathbf{G}\vec{A}$ (which follows from Eq. \ref{eq:GVHV}), we find 
\begin{equation}
    \frac{d\vec{A}}{dt} = [\iu \mathbf{G}(t) - \mathbf{V}^{-1}(t) \dot{\mathbf{V}}(t)] \vec{A}(t) \equiv \mathbf{W}(t) \vec{A}(t).
    \label{eq:dAdt}
\end{equation}
The above equation is exact, but still involves highly oscillatory complex mode amplitudes. To make further progress, we note that Eq. (\ref{eq:dAdt}) yields the evolution equation for $|A_{\alpha}(t)|$:
\begin{equation}
\frac{d}{dt} |A_{\alpha}| = \frac{1}{|A_{\alpha}|} \mathrm{Re}\Big(\sum_{\beta} A_{\alpha}^* W_{\alpha \beta} A_{\beta}\Big).
\label{eq:dabsAdt}
\end{equation}
In general, $A_{\alpha}^*$ and $A_{\beta}$ contain fast varying phases. We now adopt the ansatz that when averaging over timescales much longer than $|\lambda_{\alpha} - \lambda_{\beta}|^{-1}$ but shorter than the orbital evolution time $|a_1/\dot{a}_1|$,
\begin{equation}
    \langle e^{\iu \phi_{\alpha \beta}} \rangle = \Big\langle \frac{A_{\alpha}^* A_{\beta}}{|A_{\alpha}||A_{\beta}|} \Big\rangle \simeq 0 ~~~~ ~~~~ ~~~~ ~~~~ (\alpha \neq \beta).
    \label{eq:phi_ab}
\end{equation}
With this ansatz, Eq. (\ref{eq:dabsAdt}) reduces to 
\begin{equation}
\frac{d}{dt} |A_{\alpha}| \simeq \mathrm{Re}(W_{\alpha \alpha}) |A_{\alpha}|,
\label{eq:dabsAdt2}
\end{equation}
and the magnitude of the planet eccentricity is given by
\begin{equation}
    \langle e_i^2 \rangle = \Big\langle |\sum_\alpha A_\alpha \mathcal{E}_{\alpha i}|^2 \Big\rangle \simeq \sum_\alpha |A_\alpha|^2 |\mathcal{E}_{\alpha i}|^2.
    \label{eq:erms}
\end{equation}
If we define $B_\alpha(t) \equiv |A_\alpha(t)| \in \Re^+$ and let $\vec{B}(t)$ be the vector with components $B_{\alpha}(t)$, Eq. (\ref{eq:dabsAdt2}) can be written in matrix form as 
 \begin{equation}
     \frac{d \vec{B}(t)}{dt} = \mathrm{Re}\left[ \mathrm{Diag}(\mathbf{W})\right] \vec{B}(t). 
     \label{eq:dBdt}
 \end{equation}
Eq. (\ref{eq:dBdt}) is much easier to solve numerically than the exact Eq. (\ref{eq:dAdt}), because both the vector $\vec{B}$ and the matrix $\mathbf{W}$ are explicit functions of $\vec{a}_1(t)$, and only vary with $t$ as $a_1(t)$ varies. In particular, we can evaluate $\mathbf{W}$ from 
\begin{equation}
    \mathbf{W} = \iu \mathbf{G} - \mathbf{V}^{-1} \left(\frac{\partial \mathbf{V}}{\partial a_1} \right) \dot{a}_1,
\end{equation}
with $\dot{a}_1$ given by Eq. (\ref{eq:dadt}). 

We solve Eq. (\ref{eq:dBdt}) combined with Eqs. (\ref{eq:dadt}) and (\ref{eq:erms}) to obtain the time evolution of the RMS eccentricity and semi-major axis. Although our formalism above does not capture the short-term oscillations in eccentricity, it is nonetheless possible to know the extent of these oscillations by computing the ``instantaneous'' maximum and minimum eccentricity. The instantaneous maximum eccentricity is given by
\begin{equation}
\mathrm{max}(e_i) = \sum_\alpha B_{\alpha} |\mathcal{E}_{\alpha i}|,
\label{eq:e_max}
\end{equation}
and the minimum eccentricity is given by
\begin{equation}
    \mathrm{min}(e_i) = \sqrt{2\langle e_i^2\rangle - [\mathrm{max}(e_i)]^2}.
    \label{eq:e_min}
\end{equation}
In the above RMS-averaged formulation, the ansatz leading to Eq. (\ref{eq:dBdt}) is equivalent to assuming that the mode amplitude evolves adiabatically as $a_1$ decreases, i.e. we assume that the cross-terms corresponding to the mixing between modes average out to zero due to their incoherent phases, and only diagonal terms remain. In reality, the assumption in Eq. (\ref{eq:phi_ab}) may not hold in the later stages of orbital decay, as certain pairs of modes may become locked in either alignment or anti-alignment depending on the configuration of eigenvectors. In practice, this turns out not to be an issue since all but one mode will have decayed away by this point, leaving the question of how to handle the cross-mode terms moot, and we have found excellent agreements across the board between the approximate RMS-averaged formulation and the exact treatment. Nonetheless, the approximation made in Eq. (\ref{eq:phi_ab}) is the main source of uncertainty in our approximate formulation and may lead to errors in edge cases when modes do not vary sufficiently rapidly relative to the orbital decay timescale.

\begin{figure}
\includegraphics[width=0.95\linewidth]{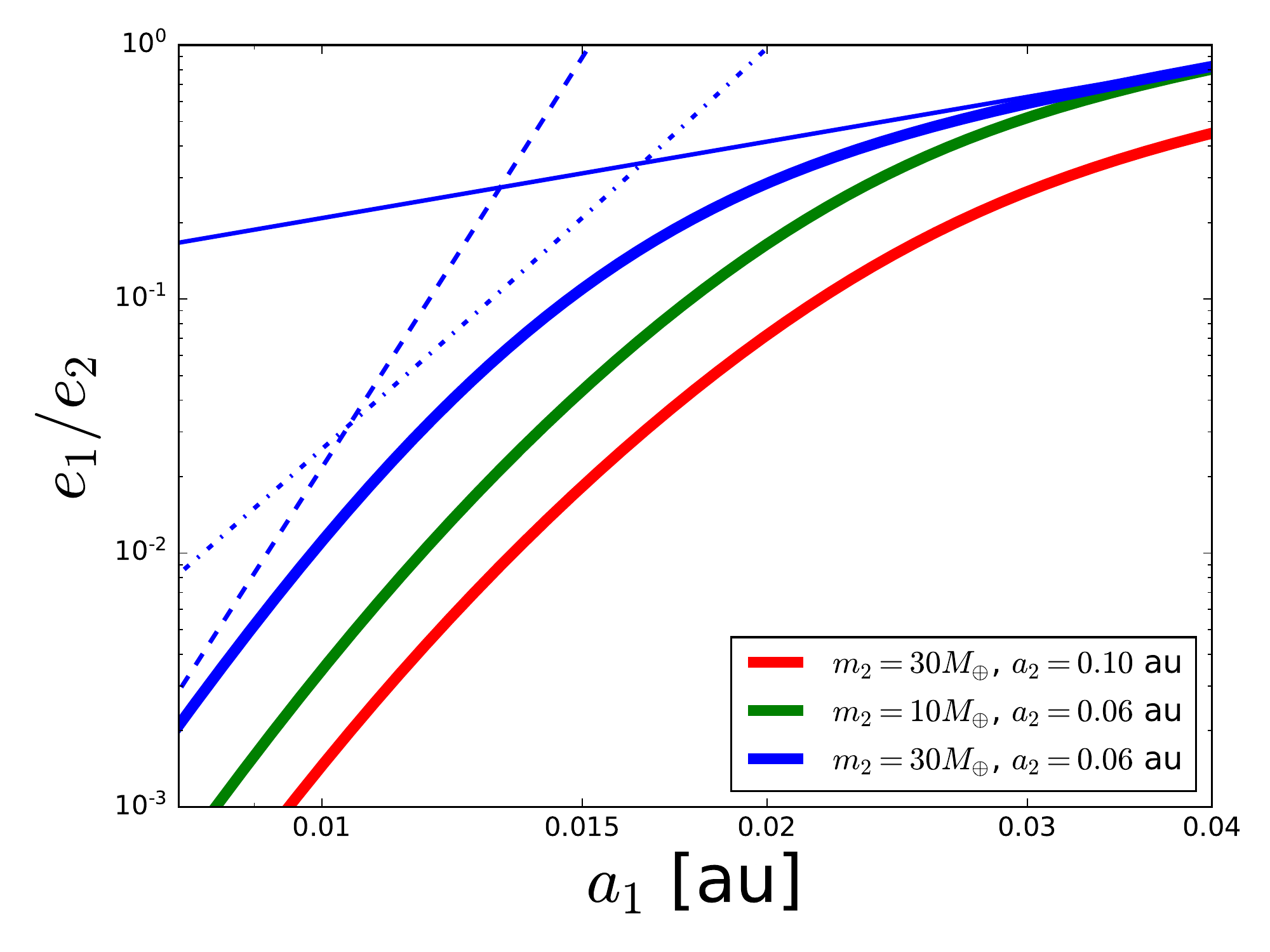}
\caption{
The value of $e_{1,\mathrm{forced}}/e_2$ (Eq. \ref{eq:e_forced}) of an inner planet with $m_1 = M_{\oplus}$ and $R_1 = R_{\oplus}$ as a function of $a_1$ for various values of $a_2$ and $m_2$. The three thick curves correspond to different values of $m_2$ and $a_2$ as labeled. 
For the blue curve, we also show its limiting cases: the thin, solid blue curve corresponds to the approximation given by $e_{1,\mathrm{forced}}/e_2 \simeq \nu_{12}/\omega_{12}$, the dash-dotted curve corresponds to $e_{1,\mathrm{forced}}/e_2 \simeq \nu_{12}/\omega_{1,\mathrm{gr}}$ and the thin dashed line corresponds to $e_{1,\mathrm{forced}}/e_2 \simeq \nu_{12}/\omega_{1,\mathrm{tide}}$.
\label{fig:fig1}
}
\end{figure}

\begin{figure}
\includegraphics[width=0.99\linewidth]{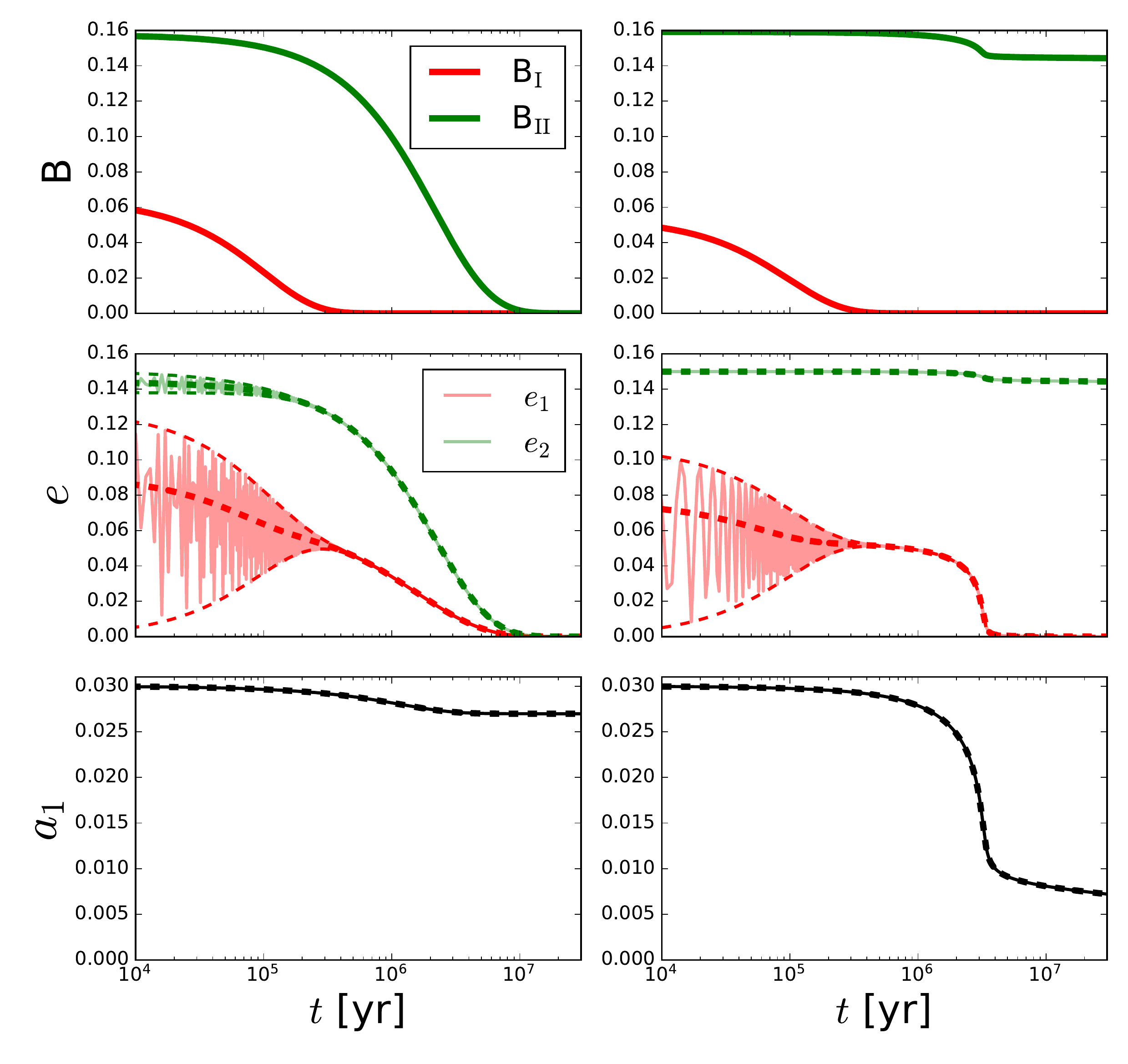}
\caption{Sample time evolution for a two-planet system with initial conditions $a_{1,0} = 0.03$ au, $a_2 = 0.08$ au and $e_{1,0} = 0$, $e_{2,0}$ = 0.15. The left panels have planet masses $m_1 = M_{\oplus}$, $m_2 = 3 M_{\oplus}$ while the right panels have $m_1 = M_{\oplus}$ and $m_2 = 30 M_{\oplus}$. Note here we adopt a large value of $\Delta t_{L,1} = 10^5 $ s in order to speed up numerical calculations. In the middle and bottom panels, the dashed curves show our approximate solution using Eq. (\ref{eq:dBdt}) while the solid curves are obtained using a direct integration of Eqs. (\ref{eq:dEdt}) and (\ref{eq:dadt}). On the left panels, the orbital decay of $m_1$ is limited by the amount of total angular momentum deficit, as the two modes decay away before substantial orbital decay can take place, whereas on the right panels, the planets have sufficient AMD to undergo substantial orbital decay, and is instead limited by the rate of orbital decay.
\label{fig:fig2}
}
\end{figure}

\section{Two-Planet Systems}
\label{sec:2.2}
\subsection{Mode Properties and General Evolution Behaviors}
We demonstrate the application of the formalism presented in section \ref{sec:2.1} by considering 2-planet proto-USP systems. In this case,  analytic expressions for the modes can be derived explicitly, providing useful insight into the more general multi-planet systems. 
The complex eigenfrequencies $\lambda_{\mathrm{I}}, ~\lambda_{\mathrm{II}}$ (see Eq. \ref{eq:gmat}) are given by
\begin{align}
\lambda_{\mathrm{I}} &= \frac{1}{2}\left(\tilde{\omega}_1 + \tilde{\omega}_2 + \sqrt{\Delta \tilde{\omega}^2 + 4\nu_{12}\nu_{21}}\right) \\
\lambda_{\mathrm{II}} &= \frac{1}{2}\left(\tilde{\omega}_1 + \tilde{\omega}_2 - \sqrt{ \Delta\tilde{\omega}^2 + 4\nu_{12}\nu_{21}}\right),
\end{align}
where $\Delta \tilde{\omega} \equiv \tilde{\omega}_1 - \tilde{\omega}_2$ (with $\tilde{\omega}_1 = \omega_1 + \iu \gamma_1$ and $\tilde{\omega}_2 \simeq \omega_2$), and the eigenvectors are
\begin{align}
\vec{\mathbf{\mathcal{E}}}_{\mathrm{I}} &= \begin{pmatrix} \Delta \tilde{\omega} + \sqrt{\Delta \tilde{\omega}^2 + 4\nu_{12}\nu_{21}} \\2\nu_{21} \end{pmatrix}, \nonumber \\ 
\vec{\mathbf{\mathcal{E}}}_{\mathrm{II}} &= \begin{pmatrix} \Delta \tilde{\omega} - \sqrt{\Delta \tilde{\omega}^2 + 4\nu_{12}\nu_{21}} \\2\nu_{21} \end{pmatrix}.
\end{align}
In general, in order for the inner planet ($m_1$) to become an USP, one requires that $L_1 \ll L_2$ (where $L_i = m_i \sqrt{GM_{\star}a_i}$ is the circular angular momentum). In this limit the above expressions simplify considerably: Since $\omega_2, ~\nu_{12}, ~\nu_{21} \ll \omega_1$ (recall that $\omega_2 \simeq \omega_{21} = \omega_{12}L_1/L_2$), the eigenfrequencies become
\begin{align}
\lambda_{\mathrm{I}} &\simeq \tilde{\omega}_1 \\
\lambda_{\mathrm{II}} &\simeq \tilde{\omega}_2 - \frac{\nu_{12}\nu_{21}}{\tilde{\omega}_1}.
\end{align}
The eigenvectors in this limit are given by
\begin{align}
\mathbf{\mathcal{E}}_{\mathrm{I}} &\simeq \begin{pmatrix} 1 \\ \frac{\nu_{21}}{\omega_1} \end{pmatrix}, ~\qquad \qquad \mathbf{\mathcal{E}}_{\mathrm{II}} \simeq \begin{pmatrix} \frac{-\nu_{12}}{\tilde{\omega}_1} \\ 1 \end{pmatrix}.
\end{align} 
Since $\nu_{21}/\omega_1 = (L_1/L_2)(\nu_{12}/\omega_1) \ll 1$, it is clear that the mode $\alpha = \mathrm{I}$ (II) is associated with the free oscillation (apsidal precession) of the inner (outer) planet. 
The damping rate rate of the two modes are given by
\begin{align}
\gamma_{\mathrm{I}} &\equiv \mathrm{Im}(\lambda_\mathrm{I}) \simeq \gamma_1 \\
\gamma_{\mathrm{II}} &\equiv \mathrm{Im}(\lambda_\mathrm{II}) \simeq \gamma_1 \frac{\nu_{12}\nu_{21}}{\omega_1^2} = \gamma_1 \left(\frac{\nu_{12}}{\omega_1}\right)^2 \frac{L_1}{L_2}.\label{eq:gamma2}
\end{align}
Clearly, the decay of eigenmode $\mathrm{II}$ is substantially supressed relative to mode $\mathrm{I}$. It is therefore safe to assume that any initial oscillation along mode $\mathrm{I}$ is quickly damped out, and the system is locked into mode $\mathrm{II}$. At this stage $e_1$ is given by the forced eccentricity:
\begin{equation}
e_1 = e_{1,\mathrm{forced}} \simeq \frac{\nu_{12}}{\omega_1} e_2 = \frac{\nu_{12} e_2}{\omega_{12} + \omega_{1,\mathrm{gr}} + \omega_{1,\mathrm{tide}}}.
\label{eq:e_forced}
\end{equation}
Figure \ref{fig:fig1} shows the ratio $e_{1,\mathrm{forced}}/e_2$ as a function of $a_1$ for several values of $m_2$ and $a_2$. We see that at large $a_1$, $\omega_1 \simeq \omega_{12}$, and we have $e_{1,\mathrm{forced}}/e_2 \simeq \nu_{12}/\omega_{12} \simeq 5a_1/4a_2$. As $a_1$ decreases, $\omega_{1,\mathrm{gr}}$ begins to dominate; in this case we have $e_{1,\mathrm{forced}}/e_2 \simeq \nu_{12}/\omega_{1,\mathrm{gr}} \propto m_2 a_1^5/a_2^4$. When $a_1$ decreases even further, $\omega_{1,\mathrm{tide}}$ becomes the most dominant term, and we have $e_{1,\mathrm{forced}}/e_2 \simeq \nu_{12}/\omega_1 \propto m_1a_1^9/a_2^4$. 

Thus, for $t \gtrsim \gamma_1^{-1}$, the orbital evolution of the inner planet is governed by (neglecting stellar tides)
\begin{equation}
\frac{\dot{a}_1}{a_1} = -2\gamma_1 e_{1,\mathrm{forced}}^2 = -2\gamma_1 \left(\frac{\nu_{12}}{\omega_1} \right)^2 e_2^2.
\end{equation}

Comparing $|\dot{a}_1/a_1|$ with $\gamma_{\mathrm{II}}$ (Eq. \ref{eq:gamma2}), we see that the system may exhibit two possible outcomes, depending on the system parameters and initial conditions: (i) for $L_{1,0}/L_2 \lesssim 2e_{2,0}^2$ (where the subscript `0' referring to the initial value), mode II does not experience significant damping (i.e. $e_2 \simeq e_{2,0}$), and the inner planet keeps undergoing orbital decay until its forced eccentricity is suppresed by GR and tides, dramatically slowing any further tidal decay. (ii) For $L_{1,0}/L_2 \gtrsim 2e_{2,0}^2$, both modes are eventually damped out, preventing further tidal decay and leaving behind two planets with circular orbits and fixed semi-major axes. 

Figure \ref{fig:fig2} shows examples of time evolution for the two cases. The solid curves are direct integrations of Eq. (\ref{eq:dEdt}) while the dashed curves utilize the approximate formulation in Eq. (\ref{eq:dBdt}); to speed up the integration of Eq. (\ref{eq:dEdt}) we have adopted an unphysical value of $\Delta t_{L,1} = 10^5$ s, corresponding to $Q_1 = 0.07$ (at $P_1 = 1$ day). Note the excellent agreement between the exact treatment and our approximation. The left three panels of Fig. \ref{fig:fig2} shows an example of case (ii) - we see that in this case the mode amplitudes $\vec{B}(t)$ decay away to zero well before the nominal orbital decay timescale $|a_1/\dot{a}_1|$. In fact, this is a general barrier to USP formation: the planets must have sufficient initial eccentricities (i.e. angular momentum deficit or AMD, see below) in order to have substantial orbital decay prior to having all the mode amplitudes dissipated away. 

The right three panels of Fig. \ref{fig:fig2} show an example of case (i). In this case the inner planet is able to undergo substantial decay to become an USP due to the greater amount of AMD stored in the more massive exterior planet, and substantial oscillation amplitude remains in mode $\mathrm{II}$ even after the inner planet has decayed to sub-day periods. We see that in this case the tidal decay is still self-limiting: As the inner planets decay further, the effects of short-range forces become important, which forces the inner planet to attain much lower $e_{1,\mathrm{forced}}$ (see Eq. \ref{eq:e_forced}) that dramatically slows down the rate of tidal decay.


\subsection{Criteria for Orbital Decay}
\label{sec:2.3}
\begin{figure}
\includegraphics[width=0.95\linewidth]{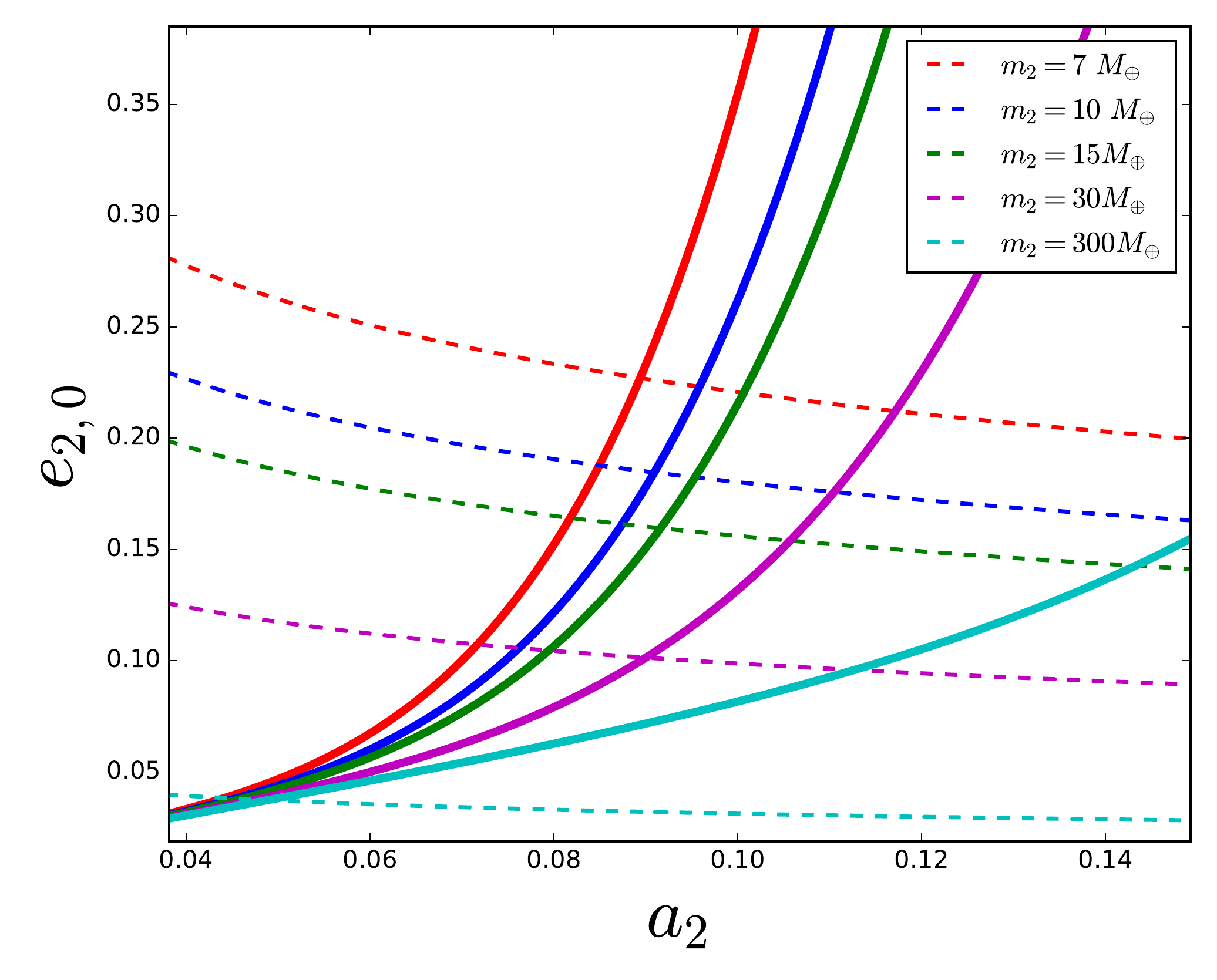}
\caption{The critical eccentricity $e_{2,0}$ needed to meet the AMD (dashed curves) and tidal decay time constraints (solid curves) for USP formation in a two-planet system is plotted as a function of the companion semi-major axis $a_2$. The inner planet has initial semi-major axis $a_{1,0} = 0.03$, mass $m_1 = M_{\oplus}$, radius $R_1 = R_{\oplus}$, and tidal lag time $\Delta t_{L,1} = 100$ s. The dashed curve are given by Eq. (\ref{eq:ecrit_1}), corresponding to $a_{1,\mathrm{min}}/a_{1,0} = \frac{1}{2}$, while the solid curves are given by Eq. (\ref{eq:ecrit_2}), corresponding to $|\dot{a}_1/a_1| \gtrsim 10^{-10}$ yr. The red, blue, green, magenta and cyan curves correspond to $m_2 = $ 7, 10, 15, 30 and 300 $M_{\oplus}$ respectively. For a given $m_2$, in order for efficient orbital decay to occur, the outer planet's initial eccentricity must be above both curves of the corresponding color.} 
\label{fig:fig4}
\end{figure}

The analysis and examples shown in Section \ref{sec:2.3} show that generally, two criteria must be met in order for the inner planet to undergo substantial tidal decay: (i) the total angular momentum deficit (AMD)\footnote{The AMD of a system is given by the difference between its total angular momentum if all planet orbits were circular and its actual angular momentum, i.e. $\mathrm{AMD} \equiv \sum_i m_i \sqrt{GM_{\star}a_i}(1 - \cos{\theta_i}\sqrt{1-e_i^2})$.} of the system $\mathrm{AMD} \simeq L_2 e_{2,0}^2/2$ must be sufficiently large to allow the inner system to undergo orbital decay before all the eccentricities are decayed away. (ii) The inner planet must have sufficiently large forced eccentricity such that the orbital decay occurs within the lifetime of the system. The first criterion arises from the conservation of the total angular momentum. Indeed, for $e_i^2 \ll 1$, the total angular momentum $L = \sum_i L_i \sqrt{1 - e_i^2} \simeq \sum_i L_i (1 - e_i^2/2)$ (with $L_i = m_i \sqrt{GM_{\star} a_i}$) evolves according to
\begin{align}
\frac{dL}{dt} = L_1 \frac{\dot{a}_1}{2a_1} - \sum_i L_i \mathrm{Re}(\dot{\mathcal{E}}_i \mathcal{E}^*),
    \label{eq:dLdt}
\end{align}
where we have omitted the orbital decay of the other planets. Substituting Eqs. (\ref{eq:dEdt}) - (\ref{eq:Ht}), and noting that $L_1 \nu_{12} = L_2 \nu_{21}$, we find that
\begin{equation}
\frac{dL}{dt} = -\frac{L_1}{2} \gamma_{\star}.
\end{equation}
Thus $L$ is constant when $\gamma_{\star}$ is negligible (which is the case until $a_1$ is already reduced to a value well below $0.02$ au by planetary tides). The semi-major axis of the inner planet decreases at the expense of the planet eccentricities, while keeping the total angular momentum constant. Assuming initially $e_{1,0} = 0$, we have
\begin{equation}
m_1 \sqrt{a_{1,0}} + m_2 \sqrt{a_2 (1-e^2_{2,0})} = m_1 \sqrt{a_1 (1 - e_1^2)} + m_2 \sqrt{a_2 (1-e^2_2)}.
\end{equation}
For a given $a_{1,0}$, $a_2$ ($= \mathrm{const.})$ and $e_{2,0}$, the minimum semi-major axis the inner planet can reach (after indefinite time) is given by 
\begin{equation}
\frac{a_{1,\mathrm{min}}}{a_{1,0}} = \left[1 - \frac{m_2 \sqrt{a_2}}{m_1 \sqrt{a_{1,0}}} \left(1 - \sqrt{1-e_{2,0}^2} \right)\right]^2 \simeq \left(1 - \frac{m_2 \sqrt{a_2}e_{2,0}^2 }{2 m_1 \sqrt{a_{1,0}}} \right)^2.
\label{eq:a1_min}
\end{equation}
Thus, the critical initial eccentricity of $m_2$ required for significant semi-major axis decay ($a_{1,\mathrm{min}}/a_{1,0} \simeq \frac{1}{2}$) is given by 
\begin{equation}
e_{2,\mathrm{crit}} = \left(2 - \sqrt{2} \right)^{1/2} \left(\frac{m_1}{m_2}\right)^{1/2} \left(\frac{a_{1,0}}{a_2}\right)^{1/4} \simeq 0.77 \left(\frac{L_{1,0}}{L_2}\right)^{1/2}.
\label{eq:ecrit_1}
\end{equation}

The second criterion pertains to the orbtial decay timescale. In order for $a_1$ to decrease significantly within $10^{10}$ yrs, the inner planet must have (see Eq. \ref{eq:adot_tide})
\begin{align}
e_{1} &\gtrsim 4.6\times10^{-3} \left(\frac{\Delta t_{L,1}}{100 s}\right)^{1/2}
\left(\frac{M_{\star}}{M_{\odot}}\right) \nonumber \\ &~~~~ \times
\left(\frac{m_1}{M_{\oplus}}\right)^{-1/2} 
\left(\frac{R_1}{R_{\oplus}}\right)^{5/2}
\left(\frac{a_1}{0.02 ~\mathrm{au}}\right)^{4}.
\label{eq:ecrit_2}
\end{align}
The above requirement is in fact overly conservative, since the rate of semi-major axis decay tends to accelerate as $a_1$ decreases until short-ranged forces become dominant (see Eq. \ref{eq:adot_tide}). Using $e_1 \simeq e_{1,\mathrm{forced}}$ (Eq. \ref{eq:eforced}), Eq. (\ref{eq:ecrit_2}) translates into another constraint on $e_{2,0}$ as a function of $a_2/a_{1,0}$. 

Thus, in the 2-planet case, the formation of USPs is limited by two constraints given by Eq. (\ref{eq:ecrit_1}) (the AMD constraint) and Eq. (\ref{eq:ecrit_2}) (with $e_1 = e_{1,\mathrm{forced}}$, the decay time constraint). These constraints are shown in Fig. \ref{fig:fig4}. The combination of these two constraints make the formation of USPs from 2-planet progenitor systems a challenging prospect; one natural way around the two barriers is to consider the effects of an additional external planet - we examine the 3-planet case in section \ref{sec:3}. 


\section{Three-Planet Systems}
\label{sec:3}
\subsection{Set-up}
\label{sec:3.1}
\begin{figure*}

\includegraphics[width=0.98\linewidth]{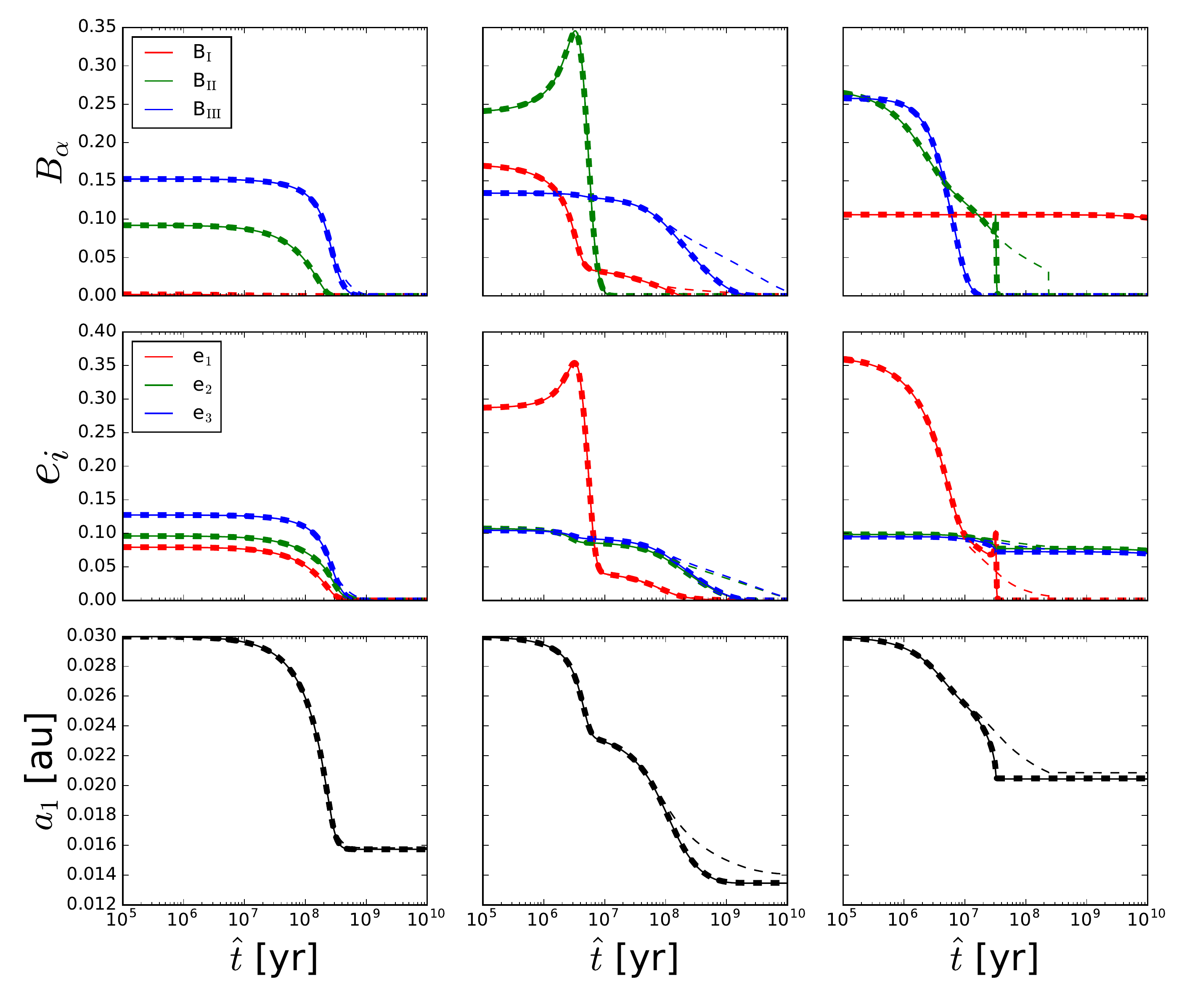}
\caption{Planet eigenmode amplitudes $B_{\alpha}$ (top), eccentricities $e_i$ (middle) and inner planet semi-major axis $a_1$ (bottom) as a function of scaled time for three different three-planet systems. Here, the time evolution is obtained from Eq. (\ref{eq:dBdt}). The thick dashed, thin solid and thin dashed lines correspond to values of $\Delta t_{L,1}$ = $10^2$, $10^3$ and $10^5$ s respectively. The time is scaled as $\hat{t} \equiv (\Delta t_{L,1} / 100 \mathrm{s}) ~t$. For the top three panels, the red, green and blue curves correspond to modes I, II and III respectively, while for the middle panels they correspond to $e_1$, $e_2$ and $e_3$. The three columns have different values of $a_2$ (0.043, 0.059 and 0.089 au respectively, from left to right), but otherwise identical parameters. The planets have masses $m_1$ = $M_{\oplus}$ and $m_2 = m_3 = 17 M_{\oplus}$ and initial semi-major axes $a_{1,0} = 0.03$ au and $a_{3} = 0.10$ au. The inner planet's radius is $R_1 = R_{\oplus}$, and stellar dissipation is neglected ($\gamma_{\star} = 0$). The initial planet eccentricities are given by $e_{1,0} = 0$ and $e_{2,0} = e_{3,0} = 0.15$, with the initial longitude of pericenter for all three planets set to $\varpi_{i,0} = 0$. 
}
\label{fig:fig8}
\end{figure*}


In this section we perform a systematic study of USP formation in 3-planet systems. The goal here is to gain physical insights on the dynamical evolution of such systems. In section \ref{sec:5} we perform population synthesis model to assess whether our model can reproduce the observed USP demographics.

We consider a 3-planet systems where the proto-USP has an initial semi-major axis $a_{1,0}$ in the range between 0.02 and 0.04 au, corresponding roughly to a period of $P_{1,0} = 1 - 3$ days. The proto-USP is assumed to have an Earth-like composition with radius $R_1 \in [1, ~1.4] ~R_{\oplus}$ with mass given by $m_1 = (R_1/R_{\oplus})^4 M_{\oplus}$, a scaling consistent with purely rocky compositions \citep{Zeng2016}. The tidal lag time $\Delta t_{L,1}$ is taken to be 100 or 1000 s, corresponding to $Q_1 = 70$ and 7 (at $P_1 = 1$ day). The outer planets have semi-major axis $a_2, a_3 \in [0.04, 0.2]$ au, masses $m_2, ~m_3 \in [3, ~30]~ M_{\oplus}$ and initial eccentricities $e_{2,0} = e_{3,0} \in [0.05, 0.3]$, and their tidal dissipation is neglected. \color{black}Note that these system parameters are chosen expediently to showcase the evolution behavior when mode mixing occurs. Some of the illustrated systems may be dynamically unstable; we discuss this issue in Sec. \ref{sec:5}.
\color{black}

As discussed in section \ref{sec:2.1}, due to the ``stiffness'' of  Eqs. (\ref{eq:dEdt}) and (\ref{eq:dadt}), we integrate Eq. (\ref{eq:dBdt}) for all our systems. For comparison purposes, we also integrate the set of systems using Eq. (\ref{eq:dEdt}), but with an enhanced value of $\Delta t_{L,1} = 10^5$ s, corresponding to an initial value of $Q_1 = 0.07$ (at $P = 1$ day); these integrations are compared to our approximate method (based on Eq. \ref{eq:dBdt}) using the same value of $\Delta t_{L,1}$. We find that our approximate formulation achieves excellent results across the entire parameter space we consider.

Note that the time evolution of systems with artificially reduced values of $Q_1$ cannot simply be considered time-scaled versions of systems with more realistic values of $Q_1$: when $Q_1 \lesssim 1$, the presence of a large imaginary component to the matrix $\mathbf{H}(t)$ substantially modifies the structure of eigenmodes, causing the inner planet eccentricity to become quantifiably different. We demonstrate this in section \ref{sec:3.2}.

\subsection{Time Evolution \& Mode Mixing}
\label{sec:3.2}
For the 3-planet case, the evolution in the framework of eigenmodes becomes considerably more complicated, and explicit analytic expressions are no longer possible. Instead, one must resort to numerical solution for the eigenfrequencies and eigenmodes. Nonetheless, the general features from the 2-planet case carry over. The mode associated with the free eccentricity oscillation of the inner-most planet tends to be damped away rapidly, while the other two modes damp on much longer timescales. However, during the evolution the eigenvalues of the three modes may cross one another, leading to substantial mode mixing. Such mixings correspond to secular resonances, causing an enhancement in the eccentricity of the inner planet and potentially speeding up its orbital decay by orders of magnitude \cite[see also][]{Hansen2015a}. This effect is most prominent when $L_1$ is much less than $L_2$ or $L_3$ \citep{Pu2018}. 

In Fig. \ref{fig:fig8}, we depict some examples of the time-evolution of the mode amplitudes $(B_{\mathrm{I}},B_{\mathrm{II}},B_{\mathrm{III}})$, the eccentricities $(e_1, ~e_2, ~e_3)$ and the inner planet semi-major axis $a_1(t)$ for three hypothetical proto-USP systems. The three systems have the same parameters and initials conditions except for different values of $a_2$. For each system, we consider three values of $\Delta t_{L,1} \in [10^2, 10^3, 10^5]$ s, corresponding to $Q_1 = [70, 7, 0.07]$ (at a period $P_1 = 1$ day). Notice that for the cases with $\Delta t_{L,1} = 10^2, ~10^3$ s, the curves (with a scaled time axis) lie right on top of each other; this shows that as long as $Q_1 \gtrsim 1$, inner planets with different values of $Q_1$ will undergo identical time evolutions if the time is scaled as $\hat{t} = t(\Delta t_{L,1}/ 100 \mathrm{s})$. On the other hand, the case with $\Delta t_{L,1} = 10^5$ s (the thin dashed lines in Fig. \ref{fig:fig8}), corresponding to an unphysical value of $Q_1 = 0.07$ (at $P = 1$ day), shows qualitatively different eccentricity and time evolutions. This demonstrates that for our parameter space, a naive approach of simply integrating Eq. (\ref{eq:dEdt}) directly with re-scaled values of $Q_1 \ll 1$ would give rise to incorrect results.

\begin{figure*}
\includegraphics[width=0.99\linewidth]{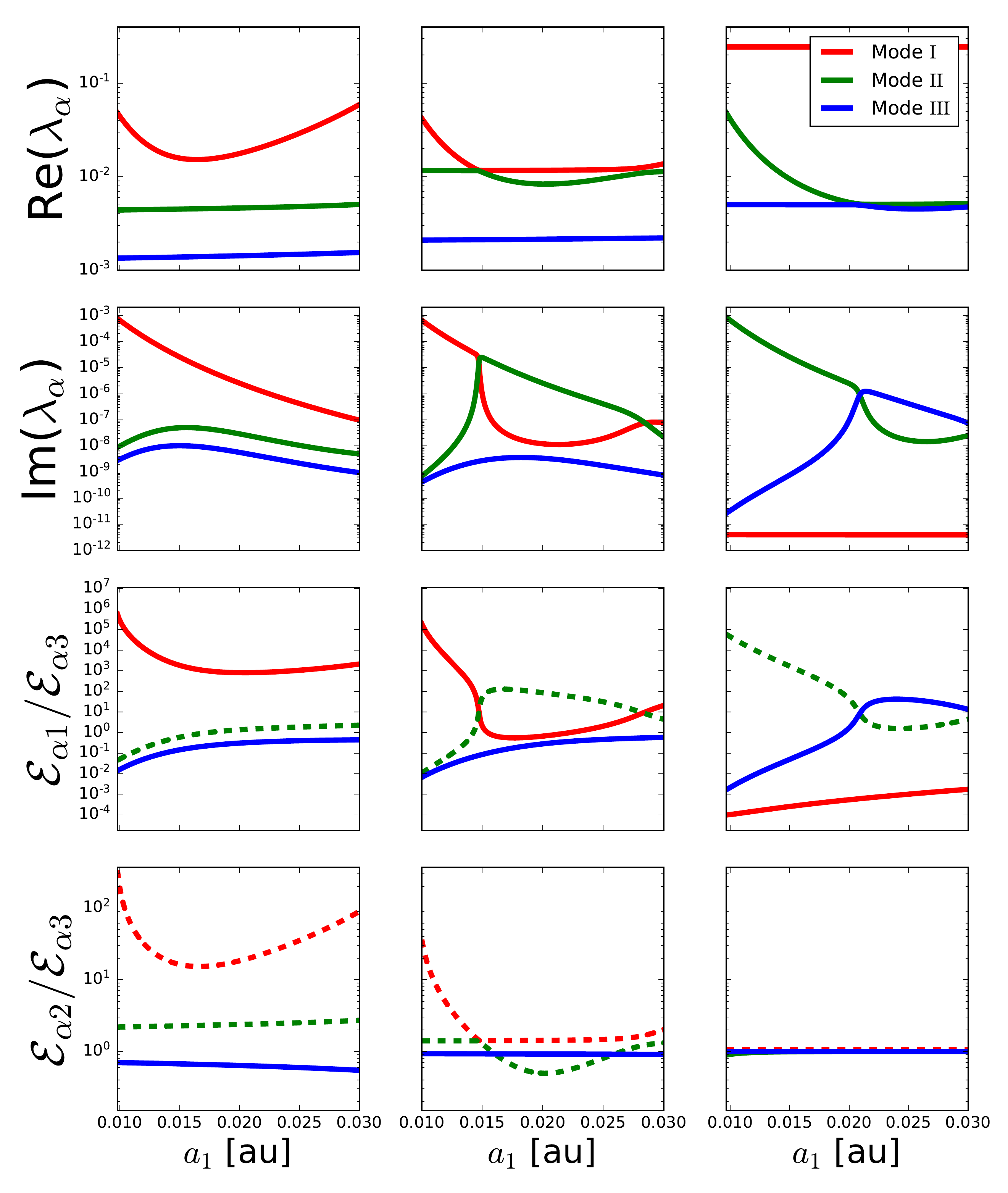}
\caption{Mode structure for the three different systems depicted in Fig. \ref{fig:fig8}. The red, green and blue curves correspond to modes I, II and III respectively. For each of the three columns, the top panel shows the real component of the eigenfrequency of the $\alpha$-th mode $\omega_{\alpha}$ as a function of inner planet semi-major axis $a_1$, while the middle-upper subpanel shows the imaginary component of the eigenvalue $\gamma_{\alpha}$. The bottom-middle subpanel shows $\mathcal{E}_{\alpha1}/\mathcal{E}_{\alpha3}$ and the bottom subpanel shows $\mathcal{E}_{\alpha2}/\mathcal{E}_{\alpha3}$. For the bottom two panels, the solid lines represent positive values while dashed lines represent negative values on the log-axis plot. The columns from left to right show three different cases for mode crossings: In the left column, the modes are well-separated and no mixing occurs; in the middle column, modes I and II show a mixing around $a_1 \approx 0.027$ au and $a_1 \approx 0.014$ au; in the right column, modes II and III cross one another at $a_1 \approx 0.022$ au.}
\label{fig:fig7}
\end{figure*}

\color{black}We now focus on the two cases with physical values of $\Delta t_{L,1}$ ($10^2$ and $10^3$ s), i.e. the thick dashed and thin solid lines of Fig. \ref{fig:fig8}. 
In Fig. \ref{fig:fig7} we show the evolution of the eigenvalues and eigenmodes for the same systems. \color{black} In the left column, the system displays no mode mixing, each mode decays independently and the proto-USP reaches a final value of $a_{1,\mathrm{f}} = 0.016$ au. In the middle column (with $a_2 = 0.059$ au), a resonance occurs between the two faster modes at $a_1 \simeq 0.027$ au (see Fig. \ref{fig:fig7}), this causes $e_1$ to increase temporarily, followed by a rapid decrease in both $e_1$ and $a_1$. After passing this resonance, the system continues to evolve, with each mode decaying independently until reaching a final value of $a_{1,\mathrm{f}} \simeq 0.014$ au\footnote{Note that a second resonance between the faster modes occurs at $a_1 \simeq 0.014 $ au; however, this does not influence the evolution because both modes have decayed to very small amplitudes by this point.}. In the right column (with $a_2 = 0.087$ au), the modes initially decay smoothly and independently of each other. A resonant mode crossing occurs at $a_1 \simeq 0.021$ au between the two slower modes. This causes $e_1$ to reverse course and increase sharply, followed by a rapid decrease in both $e_1$ and $a_1$. The inner planet reaches a final $a_{1,\mathrm{f}} \simeq 0.021$ au.

\begin{figure*}
\includegraphics[width=0.89\linewidth]{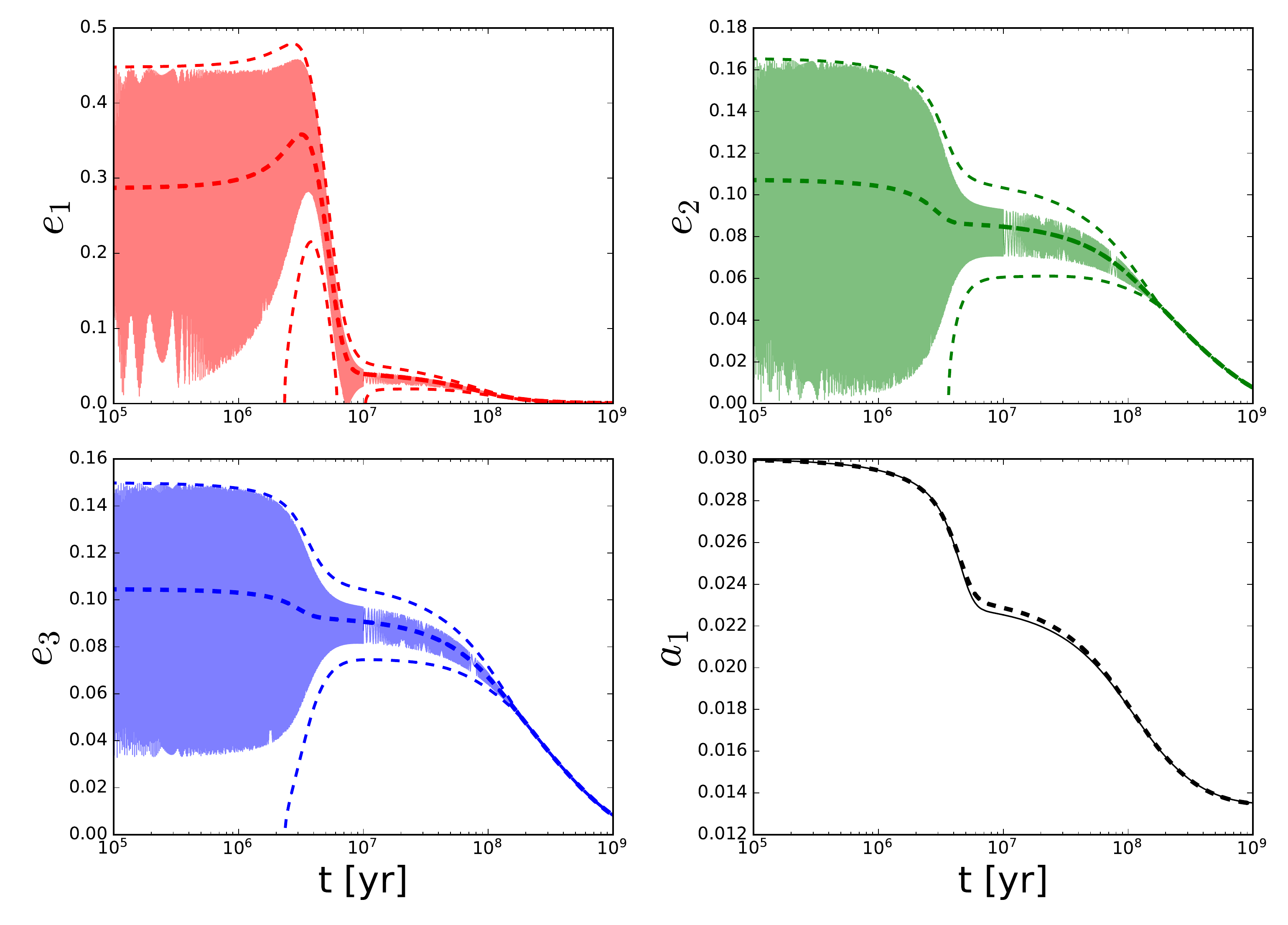}
\caption{Eccentricities $e_i$ and semi-major axis of the innermost planet $a_1$ as a function of time for the three-planet system corresponding to the right panels of Fig. \ref{fig:fig8} - \ref{fig:fig7}, with $\Delta t_{L,1} = 100$ s; the dashed curves represent the results of our approximate method based on Eq. (\ref{eq:dBdt}) while the solid lines are the results of a direct integration of Eq. (\ref{eq:dEdt}). The thick dashed curve show the values of $\langle e_i^2 \rangle^{1/2}$ (Eq. \ref{eq:erms}) while the two thin dashed lines show the maximum and minimum eccentricities given by Eq. (\ref{eq:e_max}) - (\ref{eq:e_min}). Note that there is some disagreement between the maximum and minimum values of our approximate method and the brute force calculation due to the fact that the system has insufficient time to reach the theoretical long-term extrema due to the rapid orbital decay.}
\label{fig:fig23}
\end{figure*}

To show that our method accurately captures the evolution of a resonant system, in Fig. \ref{fig:fig23} we show a comparison of the right panels of Fig. \ref{fig:fig8} - \ref{fig:fig7} with the result obtained by a brute force computation using the exact Eq. (\ref{eq:dEdt}), a result which took 5 days to complete on a Ryzen 1700 processor. Overall, there is an excellent agreement between our approximate results and the results obtained by the brute-force integration of Eq. (\ref{eq:dEdt}).

Although the presence of the secular resonance can help to speed up tidal evolution of proto-USP systems, it is not a necessary condition to form USPs. In the next section, we discuss the conditions under which USPs may form in 3-planet systems.

\subsection{Criteria for Orbital Decay}
\label{sec:3.3}
As in the 2-planet case (section \ref{sec:2.2}), the formation of USPs is constrained by two factors: (i) The amount of the total AMD to sustain the orbital decay, and (ii) the amount of forced eccentricity of the proto-USP in order to have tidal decay occur within the lifetime of the system. 
Angular momentum conservation implies that the minimum semi-major axis that can be attained by the inner planet is (cf. Eq. \ref{eq:a1_min})
\begin{equation}
    \left(\frac{a_{1,\mathrm{min}}}{a_{1,0}}\right) \simeq \left(1 - \frac{ \sum_{i \ge 2} \sqrt{a_i} m_i e_{i,0}^2}{2 m_1 \sqrt{a_{1,0}}}\right)^2.
\label{eq:a1_min3pl}
\end{equation}
Therefore, to achieve $a_{1,\mathrm{min}}/a_{1,0} \lesssim \frac{1}{2}$, one requires (assuming $e_{2,0} \sim e_{3,0}$)
\begin{equation}
    e_{2,0} \sim e_{3,0} \gtrsim 0.77 \left(\frac{m_1 \sqrt{a_{1,0}}}{\sum_{i \ge 2} \sqrt{a_i} m_i } \right)^{1/2}.
    \label{eq:ecrit_3pl}
\end{equation}
At the same time, analogous to the 2-planet case, to have efficient orbital decay within the lifetime of the system, Eq. (\ref{eq:ecrit_2}) must be satisfied. In the case of 3 planets, the inner planet forced eccentricity can no longer be expressed in a simple expression; one must solve numerically the eigenvalues and eigenvectors; the forced eccentricity can be obtained from the amplitudes of the two slower decaying modes
\begin{equation}
    e_{1,\mathrm{forced}} = \left( \sum_{\alpha \ge {\mathrm{II}}} |A_{\mathrm{\alpha}}| |\mathcal{E}_{\alpha1}| \right)^{1/2}.
    \label{eq:eforced_3pl}
\end{equation}
Since $A_{\alpha}$ is determined from the initial values of $e_{2,0}$ and $e_{3,0}$, the constraint on the inner planet eccentricity corresponds to a constraint on the external planet eccentricities. 
In the limit that $L_3 \gg L_2, ~L_1$, an approximate expression for the forced eccentricity is given by \citep[see][]{Pu2018}
\begin{equation}
    e_1 \simeq e_{1,\mathrm{forced}} = \left( \frac{\nu_{12}\omega_2 + \nu_{12}\nu_{23}} {\omega_1 \omega_2 - \nu_{12}\nu_{21}} \right) e_3. \label{eq:eforced}
\end{equation}
The above equation is more accurate when the planets are spaced evenly and well-separated, and does not fully capture the resonant mode crossings. In general, $e_{1,\mathrm{forced}}$ tends to be greater than given by the expression above, due to the contribution of other modes and aforementioned resonances. 

\begin{figure}
\includegraphics[width=1.00\linewidth]{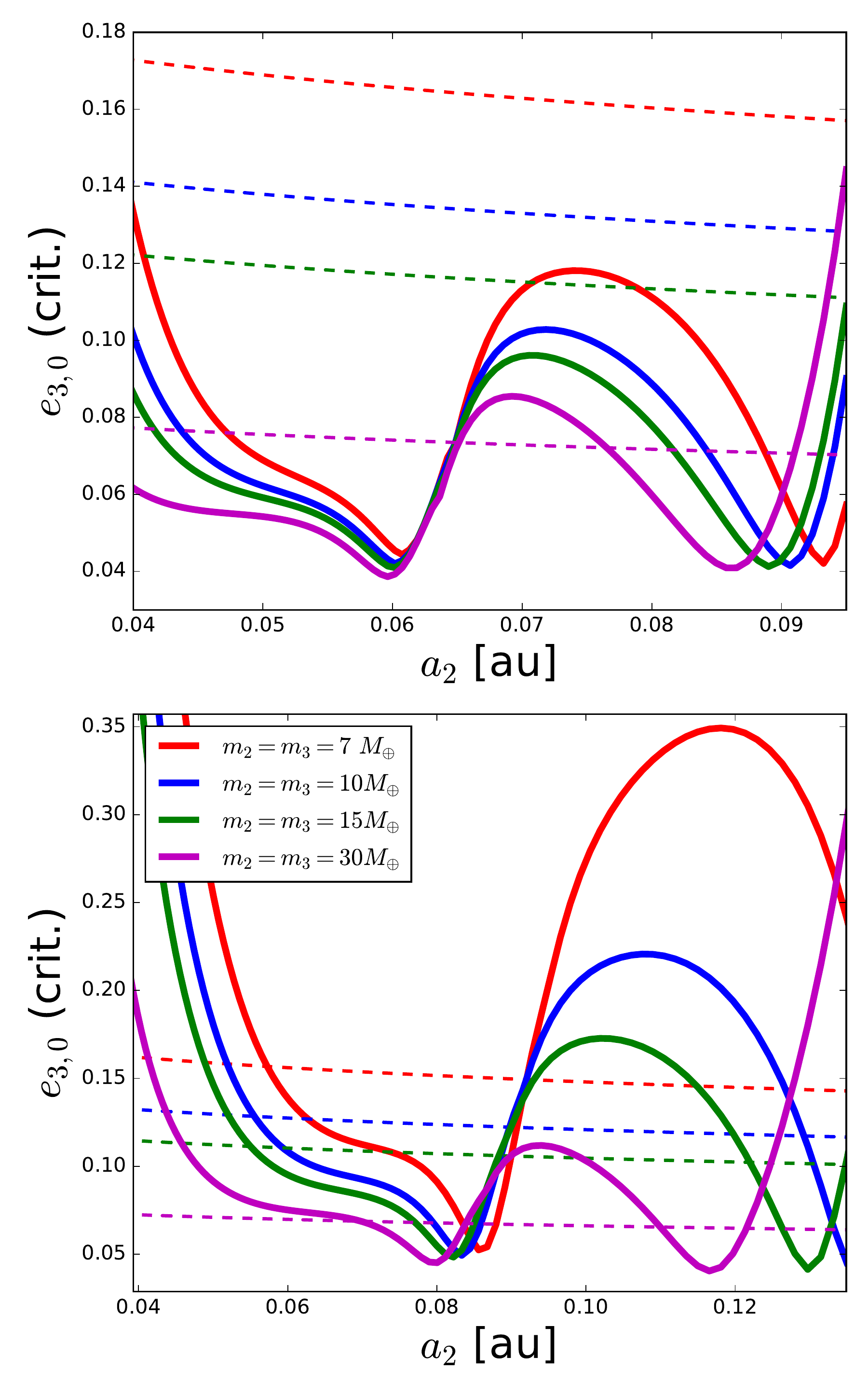}
\caption{Similar to Fig. \ref{fig:fig4}, except here the system has 3 planets. The semi-major axis of the 3rd planet is fixed at $a_3 = 0.10$ au in the top panel, and $a_3 = 0.15$ au in the bottom panel. The dashed curves (AMD constraint) are given by Eq. (\ref{eq:ecrit_3pl}) while the solid curves (decay time constraint) are given by Eqs. (\ref{eq:ecrit_2}) and (\ref{eq:eforced_3pl}), with the eigenvectors being solved numerically and assuming that $e_{2,0} = e_{3,0}$. The two dips in the solid curves correspond to the two resonant mode crossings discussed in section \ref{sec:3.2}. For a given $m_2 = m_3$, in order for efficient orbital decay to occur, the outer planet's initial eccentricities must be above both curves of the corresponding color. Note that some values of $a_2$ may result in dynamically unstable systems.
}
\label{fig:figX}
\end{figure}

In Fig. \ref{fig:figX}, we show the two constraints for USP formation in three-planet systems; this figure is analogous to Fig. \ref{fig:fig4}, except with the addition of a third planet (with $m_3 = m_2$ and fixed $a_3$). We find two important differences between the constraints for two-planet systems (see Fig. \ref{fig:fig2}) and three-planet systems: (i) {\it ceteris paribus}, the presence of an additional planet lowers the eccentricity values ($e_{2,0}, e_{3,0}$) required to meet the AMD constraint; (ii) the decay time constraint can be met by a larger set of values of $a_2$ and $e_{3,0}$, since the presence of two secular resonances makes it possible for $e_{1,\mathrm{forced}}$ to be large even for smaller values of $e_{2,0}$ and $e_{3,0}$.

\begin{figure}
\includegraphics[width=0.99\linewidth]{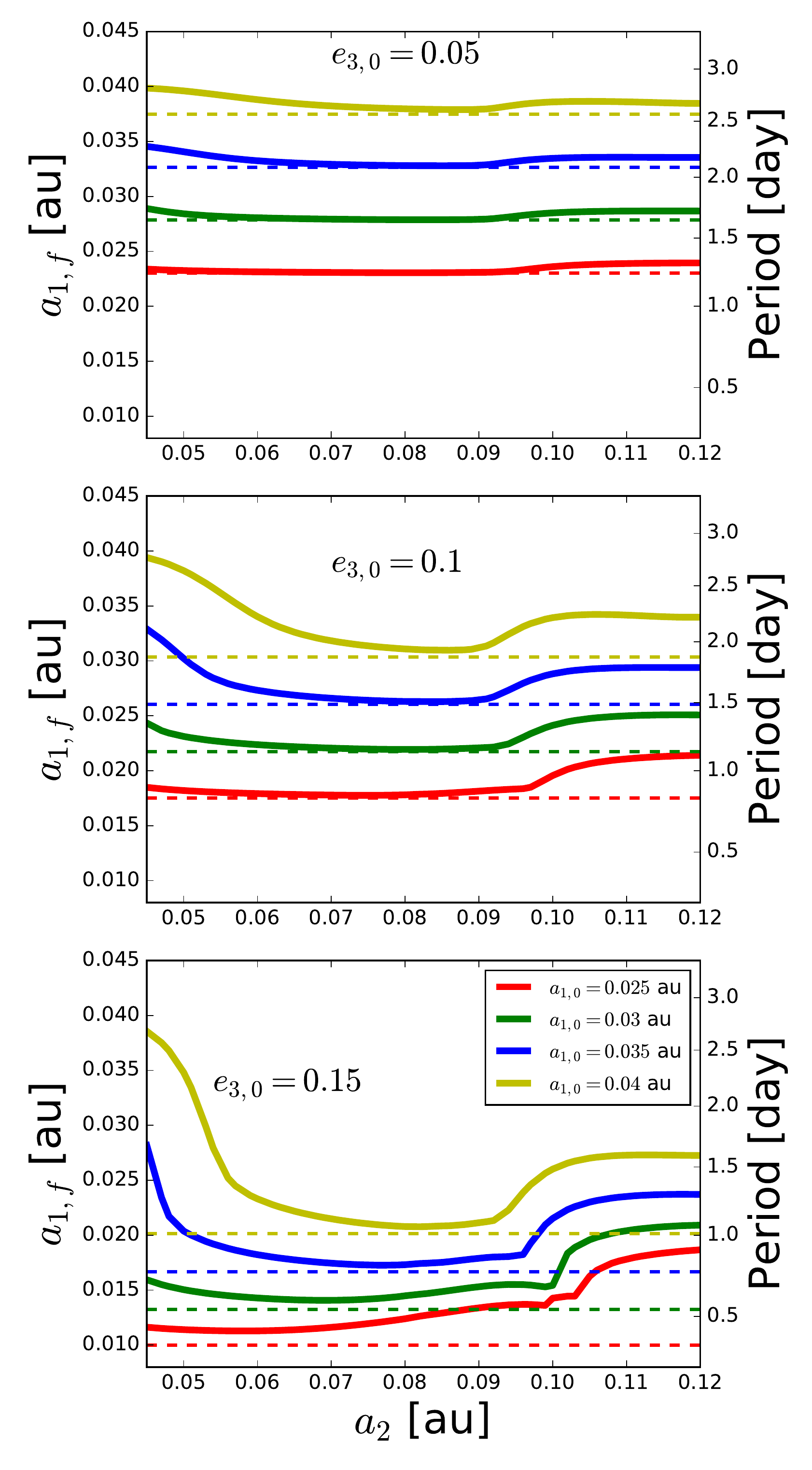}
\caption{The final value of $a_{1,\mathrm{f}}$ after 10 Gyr of evolution in a three-planet system (solid curves) and its theoretical minimum dictated by the AMD constraint (dashed curves, Eq. \ref{eq:a1_min3pl}), plotted as a function of $a_2$. The inner planet has mass $m_1 = M_{\oplus}$, radius $R_1 = R_{\oplus}$ and tidal lag time $\Delta t_{L,1} = 100$ s, while the outer planets have $m_2 = m_3 = 13 M_{\oplus}$. The three panels correspond to different initial values of $e_{2,0} = e_{3,0}$ (with initially aligned pericenters), as indicated. For each panel, the semi-major axis of the outer planet is fixed at $a_3 = 0.15$ au, while $a_2$ is varied; the red, blue, green and yellow curves correspond to $a_{1,0} = $ 0.025, 0.03, 0.035 and 0.04 au respectively. Regions where the solid curves lie on top of the dashed curves indicate the system is AMD-constrained, while regions where the solid curve is well separated from the dashed curve correspond to tidal decay time-constrained systems.}
\label{fig:fig_decay}
\end{figure}

In general, for three-planet systems, the AMD constraint is more stringent than the decay time constraint. To illustrate this, we show the final value of $a_{1,\mathrm{f}}$ reached after 10 Gyr of evolution as a function of $a_2$ in Fig. \ref{fig:fig_decay} for three-planet systems with varying initial values of $a_{1,0}$, with the planet masses, $a_3$ and $e_{2,0} = e_{3,0}$ fixed. The dashed curves in Fig. \ref{fig:fig_decay} correspond to the minimum possible value of $a_{1,\mathrm{f}}$ given by the AMD constraint, while the solid curves are their actual values at the end of the evolution. For systems with $e_{3,0} \lesssim 0.1$, the solid curve comes very close to the dashed curve, indicating that the orbital decay of $a_1$ is being stalled by a lack of AMD. For systems with $e_{3,0} \gtrsim 0.15$, the orbital decay instead becomes time-limited. 

The fact that USP production is more constrained by AMD has certain observational implications. One would expect USPs to be systematically lower in mass, as lower-mass inner planets are more likely to meet the AMD constraint (see Eq. \ref{eq:ecrit_3pl}). At the same time, we expect the external companions of USPs to have systematically larger masses, although giant planet companions are not required. To generate USPs efficiently, we also require the primordial planet eccentricities to be $e_{2,0} \sim e_{3,0} \gtrsim 0.1$, although their final values can be much lower due to tidal dissipation. The observational implications are explored in more detail in section \ref{sec:5}, where we develop a population model for USP generation. 

In this section. we have explored USP formation from three-planet systems. At first glance, there is a tension between USP generation from multi-planet systems and the fact that observed USPs have a dearth of exterior transiting companions compared with their non-USP counterparts. This {\it prima facie} contradiction can be rectified when we consider the mutual inclination evolution of USP-forming systems, in section \ref{sec:4}.

\section{Inclination Evolution}
\label{sec:4}
\color{black}We are interested in the inclination evolution of the proto-USP system because the evolution of the mutual inclination of planets determine the extent to which USPs will transit simultaneously with their companions, a quantity that can be observationally constrained (see Sec. \ref{sec:5}). As we shall demonstrate in this section, there exists a secular mutual inclination `resonances' that roughly coincides with the secular eccentricity resonance; systems that result in large excitations in eccentricity (and therefore forming USPs) should also expect large excitations in mutual inclination.  
\color{black}

The inclination evolution of the proto-USP system proceeds in an analogous fashion as the eccentricity evolution. We define the complex variable $\mathcal{I}_j = \theta_j \exp(\iu \Omega_j)$ for each planet, where $\theta_j$ is the orbital inclination (relative to the initial orbital plane) and $\Omega_j$ is the longitude of the ascending node. The mutual inclination $\theta_{ij}$ between planets $i$ and $j$ is given by
\begin{equation}
    \theta_{ij}(t) = |\mathcal{I}_i(t) - \mathcal{I}_j(t)|.
\end{equation}
For convenience we define the inclination of the N-planet system as
\begin{equation}
    \vec{\mathcal{I}} = \begin{pmatrix} \mathcal{I}_1 \\ \mathcal{I}_2 \\ \vdots  \end{pmatrix}.
\end{equation}
For small inclinations (i.e. $\theta_i \ll 1$) the time evolution of $\vec{\mathcal{I}}$ is governed by
\begin{align}
\frac{d}{dt} \vec{\mathcal{I}}(t) &= \iu \mathbf{H'}(t) \vec{\mathcal{I}}(t) + \iu \vec{\omega}_{\star} \mathcal{I}_{\star},
\label{eq:dIdt}
\end{align}
where $\mathcal{I}_{\star}$ is the complex obliquity of stellar spin. In Eq. (\ref{eq:dIdt}), the first term in the RHS is due to secular planet-planet interactions, while the second term accounts for the nodal precession driven by the stellar spin; the vector $\vec{\omega}_{\star}$ is given by
\begin{equation}
\vec{\omega}_{\star} = \begin{pmatrix} \omega_{1\star} \\ \omega_{2\star} \\ \vdots \\ \end{pmatrix}.
\end{equation}
The $N \times N$ matrix $\mathbf{H'}(t)$ is given by
\begin{equation}
\mathbf{H'}(t) =  \begin{pmatrix}
  -\omega'_{1} & \omega_{12} & \cdots & \omega_{1N}  \\
  \omega_{21} & -\omega'_{2} & \cdots & \omega_{2N}   \\
  \vdots  & \vdots  & \ddots & \vdots \\
  \omega_{N1} & \omega_{N2} & \cdots & -\omega'_{N}
 \end{pmatrix}
\end{equation}
where $\omega_{ij}$ is given by Eq. (\ref{eq:wjk}), and
\begin{equation}
\omega'_i = \sum_{j \neq i} \omega_{ij} + \omega_{i\star}.
\end{equation}
We also need to account for the evolution of stellar spin, governed by
\begin{equation}
    \frac{d\mathcal{I}_{\star}}{dt} = \iu \sum_j \omega_{\star j} \mathcal{I}_{j}.
\end{equation}
The nodal precession rate of the $i$-th planet driven by the stellar spin-induced quadrupole is
\begin{align}
\omega_{i \star} &= \frac{3k_{q\star}}{2k_{\star}} \left(\frac{m_i}{M_{\star}}\right) \left(\frac{R_{\star}}{a_i}\right) \left(\frac{S_{\star}}{L_i} \right) \Omega_{\star} 
= 2.7\times 10^{-5} 
\left(\frac{k_{q\star}}{0.01} \right) \nonumber
\\ &~ ~ ~ ~\times
\left(\frac{a_i}{0.02 \mathrm{au}} \right)^{-7/2}
\left(\frac{M_{\star}}{M_{\odot}} \right)^{-1/2}
\left(\frac{R_{\star}}{R_{\odot}} \right)^5
\left(\frac{P_{\star}}{30 \mathrm{days}} \right)^{-2} ~\mathrm{yr}^{-1},
\end{align}
where $M_\star, ~R_\star$ and $\Omega_\star = 2\pi/P_{\star}$ are the stellar mass, radius and angular rotation frequency respectively. The constants $k_\star$ and $k_{q\star}$ are defined through the star's moment of inertia and quadrupole moment: $I_{\star3} = k_\star M_\star R_{\star}^2$ and $I_{\star 3} - I_{\star 1} = k_{q\star} \hat{\Omega}^2_{\star} M_{\star} R_{\star}^2$ where $\hat{\Omega}^2_\star = \Omega_{\star}(GM_{\star}/R_{\star}^3)^{-1/2}$. Typical values for solar type stars are $k_\star \simeq 0.06$ and $k_{q\star} \simeq 0.01$ \citep[e.g.][]{Lai2018}.
The ratio of the stellar spin angular momentum $S_{\star} = I_{\star3} \Omega_{\star}$ to the orbital angular momentum of the $i$-th planet $L_i$ is
\begin{equation}
\frac{S_\star}{L_i} = 35 
\left(\frac{k_{\star}}{0.06} \right)
\left(\frac{m_i}{M_{\oplus}} \right)^{-1}
\left(\frac{a_i}{0.02 \mathrm{au}} \right)^{-1/2}
\left(\frac{M_{\star}}{M_{\odot}} \right)^{1/2}
\left(\frac{R_{\star}}{R_{\odot}} \right)^2
\left(\frac{P_{\star}}{30 \mathrm{days}} \right)^{-1},
\end{equation}
The precession rate of the stellar spin driven by the $i$-th planet is
\begin{align}
\omega_{\star i} &= \omega_{i \star} \frac{L_i}{S_\star}
= 7.7\times 10^{-7} 
\left(\frac{6k_{q\star}}{k_{\star}} \right)
\left(\frac{m_i}{M_{\oplus}} \right)
\nonumber \\ &~ ~ ~ ~\times
\left(\frac{a_i}{0.02 \mathrm{au}} \right)^{-3}
\left(\frac{M_{\star}}{M_{\odot}} \right)^{-1}
\left(\frac{R_{\star}}{R_{\odot}} \right)^3
\left(\frac{P_{\star}}{30 \mathrm{days}} \right)^{-1} ~\mathrm{yr}^{-1} .
\end{align}
For small planets ($m_1 \ll 35 M_{\oplus}$) at $P_1 \sim$ 1 day periods, the stellar spin angular momentum is much greater than the orbital angular momentum. Thus we can assume the stellar spin axis constantly points towards the $\hat{z}$-axis, i.e. $\mathcal{I}_{\star} \simeq 0$.

The rotation rate of the star $\Omega_\star$ decreases due to magnetic braking. According to \cite{Skumanich1972}, $\dot{\Omega}_\star \propto -\Omega_{\star}^3$, so that the time evolution of the spin rate is given by
\begin{equation}
    \Omega_\star = \frac{\Omega_{\star,0}}{\sqrt{1 + \alpha_{\mathrm{MB}} \Omega_{\star,0}^2 t}},
    \label{eq:skumanich}
\end{equation}
where $\Omega_{\star,0}$ is the initial spin rate, and  $\alpha_{\mathrm{MB}}$ is a constant, calibrated such that the rotation period reaches $\sim 30$ days at an age $\sim 5$ Gyr. For this section, we adopt a constant value of $\Omega_{\star}$ in lieu of the Skumanich law to better control for the effect of stellar spin; the effect of a time-dependent stellar spin period is left for section \ref{sec:5}.

The above equations, coupled with the time-evolution of the planet eccentricities $e_i$ and inner planet semi-major axis $a_1$ fully describes the inclination evolution of the system in the linear regime ($e_i, \theta_i \ll 1$). Analogous to the case of eccentricity evolution, Eq. (\ref{eq:dIdt}) involves terms that oscillate rapidly compared with the timescale of orbital decay, leading to a `stiff' set of equations that resists brute-force simulations. In section \ref{sec:4.1} we address this issue by recasting the problem in the framework of eigenmodes.

\subsection{Inclination Evolution in the Framework of Eigenmodes}
\label{sec:4.1}


\begin{figure}
\includegraphics[width=1.05\linewidth]{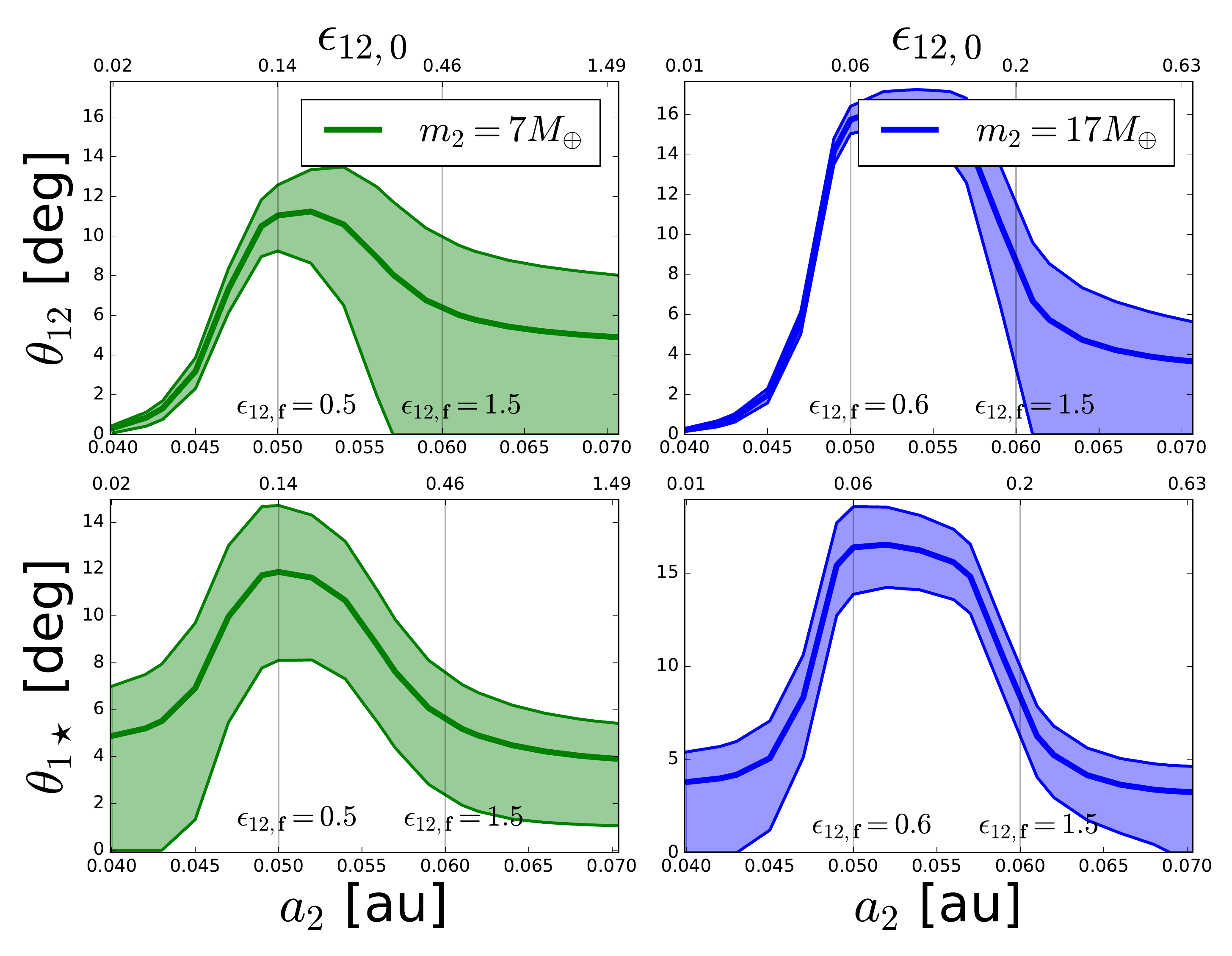}
\caption{Final values of the inner planet mutual inclination $ \theta_{12,\mathrm{f}}$ (top) and the spin-orbit angle $\theta_{1\star, \mathrm{f}}$ (bottom) as a function of $a_2$ for a three-planet system with $m_1 = M_{\oplus}$, $m_2 = 7 M_{\oplus}$ (green curves, left panels) or $17 M_{\oplus}$ (blue curves, right panels) and $m_3 = 17 M_{\oplus}$.  The initial semi-major axes of the planets are $a_{1,0} = 0.03$ au and $a_3 = 0.10$ au, while $a_2$ is  varied between 0.04 to 0.07 au. The stellar rotation period is set to $P_\star = 30$ days. The initial eccentricities are $e_{1,0} = 0$ and $e_{2,0} = e_{3,0} = 0.15$, such that the inner planet reaches $a_{1,\mathrm{f}} \approx 0.017$ au after 10 Gyr of tidal decay (note that $a_{1,\mathrm{f}}$ can be slightly different for different values of $a_2$). The initial inclination is given by $\mathcal{I}_{1,0} =  \mathcal{I}_{2,0} = 0$ and $\mathcal{I}_{3,0} = 0.075$. The bolded curves are the final RMS values given by Eqs. (\ref{eq:dirms}) and (\ref{eq:irms}) for the top and bottom panels respectively, while the two thin curves are their ``instantaneous'' maximum and minimum values given by Eqs. (\ref{eq:imax}) - (\ref{eq:imin}) and (\ref{eq:thetajk_max}) - (\ref{eq:thetajk_min}) for the bottom and top panels respectively. The top axis of each panel shows the initial value of $\epsilon_{12,0}$ (Eq. \ref{eq:ep12}). The two thin vertical lines show the the values of $a_2$ (or $\epsilon_{12,0}$) that lead to specific values of $\epsilon_{12,\mathrm{f}}$ (as indicated).
}
\label{fig:fig_theta}
\end{figure}

\begin{figure}
\includegraphics[width=1.02\linewidth]{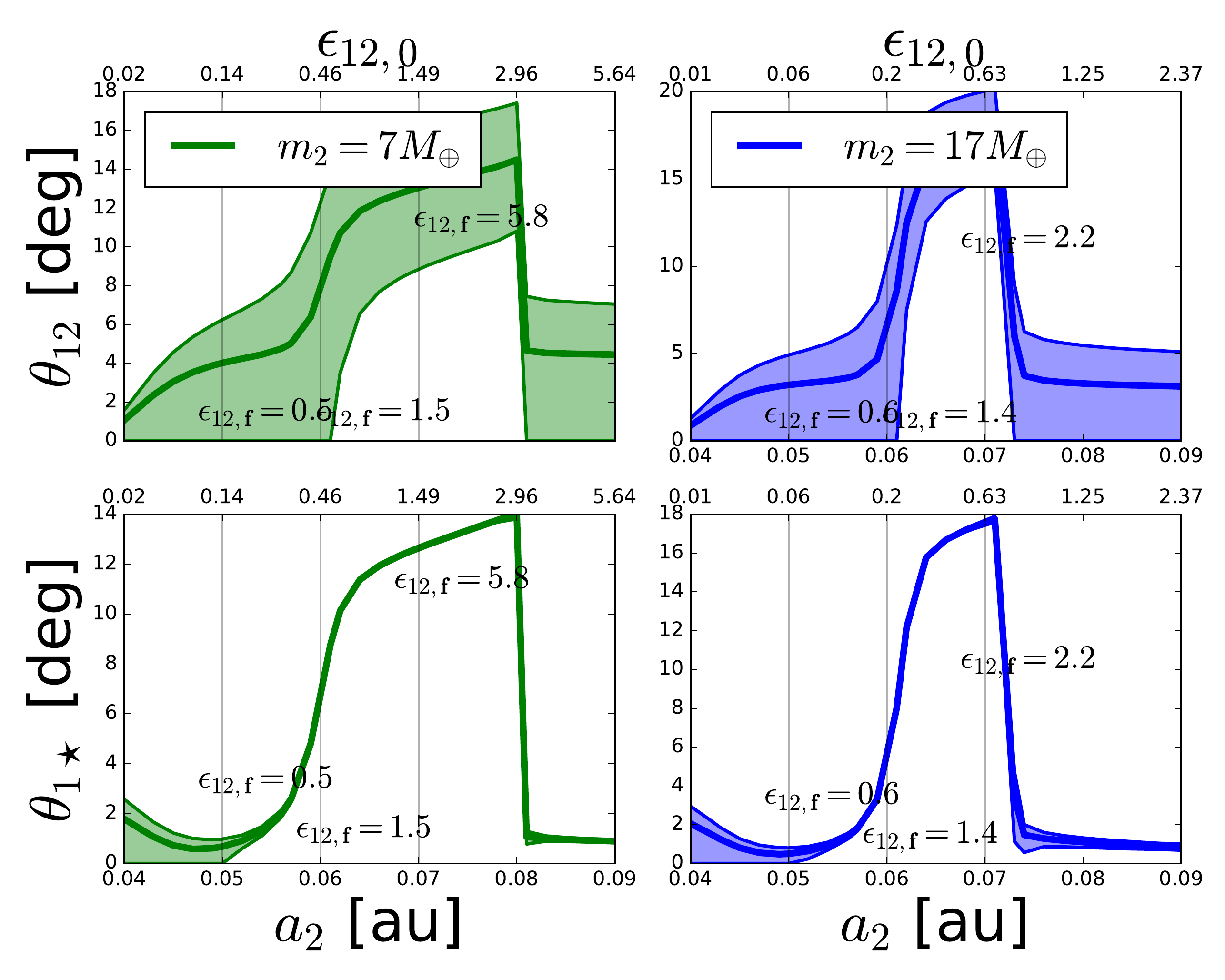}
\caption{Same as Fig. \ref{fig:fig_theta}, except with the stellar spin period fixed at $P_{\star} = 1$ day.}
\label{fig:fig_Pstar}
\end{figure}

In the discussion below we assume that the stellar spin axis is always along the $\hat{z}$-axis. In this case Eq. (\ref{eq:dIdt}) simplifies to
\begin{equation}
    \frac{d \vec{\mathcal{I}}}{dt} \simeq \iu \mathbf{H'} \vec{\mathcal{I}}.
\end{equation}
We define the eigenvalue $\lambda'_{\alpha}$ and eigenvector $\mathcal{I}_{\alpha}$ with modes denoted using Roman Numerals ($\alpha \in [\mathrm{I, ~II, ~III...}]$) of the system as
\begin{equation}
    \mathbf{H'}\mathcal{I}_{\alpha} = \lambda'_{\alpha} \vec{\mathcal{I}}_\alpha,
\end{equation}
where we have define the vector eccentricity vector $\vec{\mathcal{I}}$ to be
\begin{equation}
    \vec{\mathcal{I}}_\alpha = \begin{pmatrix} \mathcal{I}_{\alpha1} \\ \mathcal{I}_{\alpha2} \\ \vdots \end{pmatrix}.
\end{equation}
As in section \ref{sec:2.1}, we introduce the matrices $\mathbf{G}'(t)$ and $\mathbf{V}'(t)$ constructed from the eigenvalues ($\lambda_{\alpha}')$ and eigenvectors ($\mathcal{I}_{\alpha})$ of $\mathbf{H}'(t)$:
\begin{equation}
\mathbf{G'}(t) = 
\mathrm{diag}(\lambda'_{\mathrm{I}}, \lambda'_{\mathrm{II}}, \hdots, \lambda'_{\mathrm{N}})
  \label{eq:gprime_mat}
\end{equation}
and
\begin{equation}
\mathbf{V}(t) = 
\begin{bmatrix}
 \vec{\mathcal{I}}_{\mathrm{I}}  & \vec{\mathcal{I}}_{\mathrm{II}}  & \hdots & \vec{\mathcal{I}}_{\mathrm{N}} 
\end{bmatrix}.
\end{equation}
The time evolution of $\vec{\mathcal{I}}$ can be written as a superposition of eigenmodes
\begin{equation}
    \vec{\mathcal{I}}(t) = \sum_\alpha^{N} C_{\alpha} \vec{\mathcal{I}}_{\alpha}  = \mathbf{V'}(t) \vec{C}(t),
\end{equation}
where $\vec{C}(t)$ is the vector of eigenmode amplitudes:
\begin{equation}
    \vec{C} \equiv \begin{pmatrix} C_{\mathrm{I}} \\ C_{\mathrm{II}} \\ \vdots \end{pmatrix}
\end{equation}
whose initial value $\vec{C}(0)$ is
\begin{equation}
    \vec{C}(0) = \mathbf{V'}^{-1}(0) ~\vec{\mathcal{I}}(0).
\end{equation}
The time evolution of $\vec{C}$ is governed by
\begin{equation}
    \frac{d\vec{C}}{dt} = [\iu \mathbf{G'}(t) - \mathbf{V'}^{-1}(t) \dot{\mathbf{V}'}(t)] \vec{C}(t) \equiv \mathbf{W'}(t) \vec{C}(t).
\end{equation}
The above equation is exact. Similar to the case of the eccentricity evolution (section \ref{sec:2.1}), we bypass the stiffness of the above equation by focusing only the evolution of $D_\alpha \equiv |C_{\alpha}|$, whose evolution is given by 
 \begin{equation}
     \frac{d \vec{D}(t)}{dt} =  -\mathrm{Diag}[\mathbf{W}'(t)] \vec{D}(t).
     \label{eq:dDdt}
 \end{equation}
Note here that $\mathbf{W}'(t)$ can also depend on the spin-down of the star, i.e.
\begin{equation}
    \mathbf{W}' = \iu \mathbf{G}' - \mathbf{V'}^{-1} \left[ \left(\frac{\partial \mathbf{V'}}{\partial a_1} \right) \dot{a}_1 +  \left(\frac{\partial \mathbf{V'}}{\partial P_\star} \right) \dot{P}_\star \right].
\end{equation}
The instantaneous RMS inclination is given by
\begin{equation}
    \langle \theta^2_i \rangle = \Big \langle |\sum_\alpha C_\alpha(t) \mathcal{I}_{\alpha i}(t) |^2 \Big \rangle = \sum_\alpha D^2_\alpha(t) |\mathcal{I}_{\alpha i}(t)|^2
    \label{eq:irms}.
\end{equation}
The ``instantaneous'' maximum inclination is given by
\begin{equation}
\mathrm{max}(\theta_{i}) \simeq \sum_\alpha D_\alpha(t) |\mathcal{I}_{i,\alpha}(t)|,
    \label{eq:imax}
\end{equation}
while the minimum inclination is 
\begin{equation}
    \mathrm{min}(\theta_{i}) = \left[2\langle \theta_i^2\rangle - \mathrm{max}(\theta_{i})^2 \right]^{1/2}.
    \label{eq:imin}
\end{equation}
Using the mode solution, we can also obtain the mutual inclination between planets. The RMS mutual inclination between planets $i$ and $j$ is given by
\begin{equation}
    \langle \theta_{ij}^2 \rangle^{1/2} = \sum_\alpha D^2_\alpha |V'_{i,\alpha} - V'_{j,\alpha}|^2,
    \label{eq:dirms}
\end{equation}
while the maximum and minimum mutual inclinations are respectively given by
\begin{align}
    \mathrm{max}(\theta_{ij}) &= \sum_\alpha D_\alpha |V'_{i,\alpha} - V'_{j,\alpha}|
    \label{eq:thetajk_max} \\
    \mathrm{min}(\theta_{ij}) &= \left[2\langle \theta_{ij}^2\rangle - \mathrm{max}(\theta_{ij})^2 \right]^{1/2}.
    \label{eq:thetajk_min}
\end{align}

\begin{figure}
\includegraphics[width=0.95\linewidth]{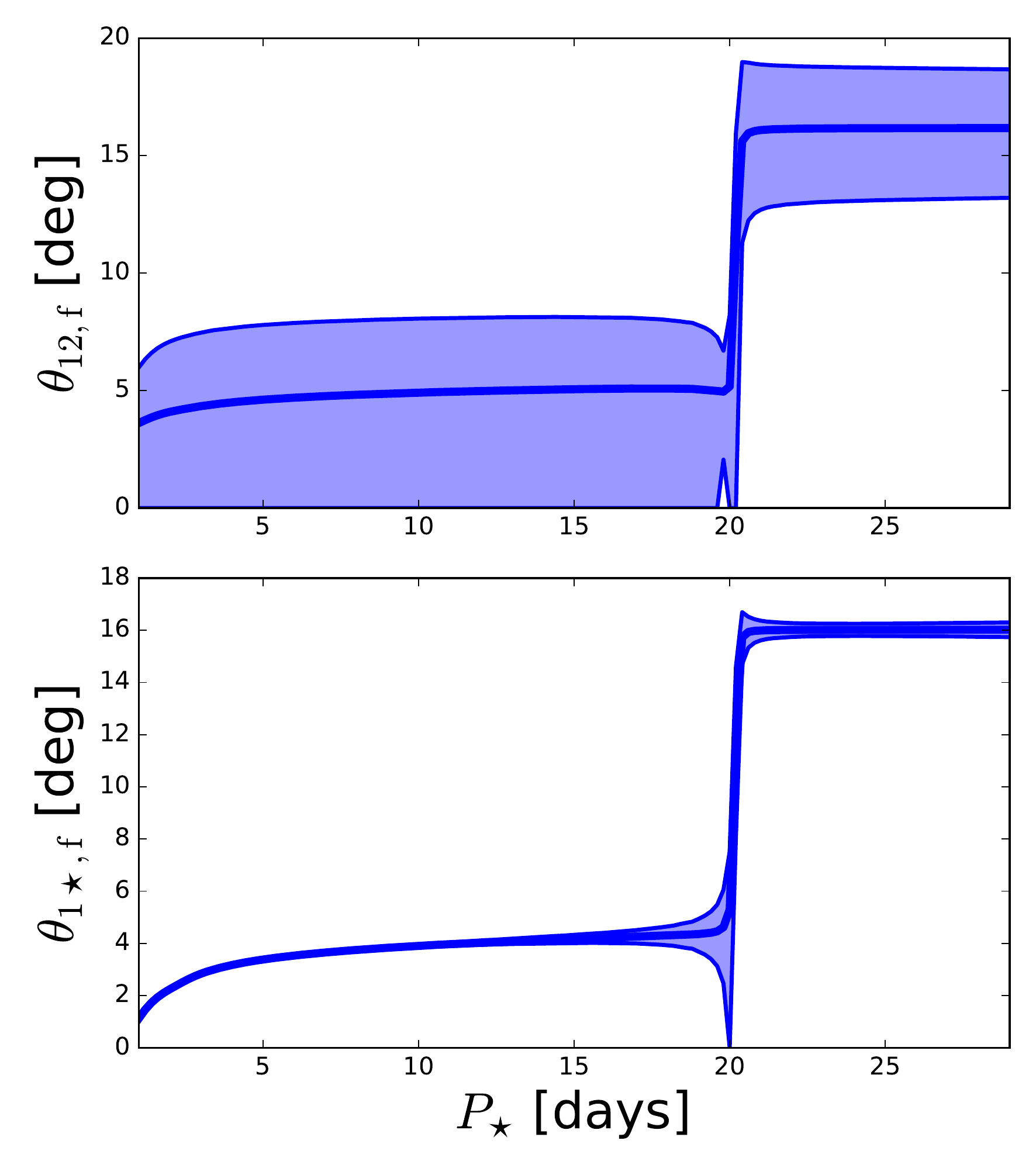}
\caption{Similar to the right panels of Fig. \ref{fig:fig_theta} (corresponding to $m_2 = 17M_{\oplus}$), except we fix the value of $a_2 = 0.055$ au, and instead vary the value of $P_{\star}$, which is fixed in time during the evolution.}
\label{fig:spin_plot}
\end{figure}

\subsection{Resonance Crossing and Mutual Inclination Excitation}
\label{sec:4.2}

Significant mutual inclinations between the inner two planets can be excited when the planet system crosses a secular inclination resonance. When spin-orbit coupling is negligible (i.e. $\omega_{i\star}$ is small), and $L_3 \gg L_1, ~L_2$, the resonance occurs when the dimensionless ``coupling parameter'' $\epsilon_{12}$, defined by
\begin{equation}
    \epsilon_{12} \equiv \frac{\omega_{23} - \omega_{13}}{\omega_{12} + \omega_{21}}
    \label{eq:ep12}
\end{equation}
is of order unity \cite[see][]{Lai2017}. As the innermost planet decays in semi-major axis, the system may transition from $\epsilon_{12} \lesssim 1$ (strong coupling between the inner two planets) to $\epsilon_{12} \gtrsim 1$ (weak coupling), crossing the resonance and generating appreciable mutual inclination $\theta_{12}$. If the orbits of $m_1$ and $m_2$ are initially co-planar, and $m_3$ is initially inclined with the inner two planets at an angle $\theta_3$, then the typical mutual inclination excited is of order \citep{Pu2018}
\begin{equation}
    \theta_{12,\mathrm{f}} \sim \theta_{3} \sqrt{\frac{L_1}{L_2}}.
\end{equation}

To illustrate the possibility of resonance, in Fig. \ref{fig:fig_theta} we show the final values of the mutual inclination between the inner two planets ($\theta_{12,\mathrm{f}}$), and the angle between the inner planet's orbit and the spin axis of its host star ($\theta_{1\star,\mathrm{f}})$\footnote{Since the stellar spin axis does not vary in our calculation, this angle is simply $\theta_{1,\mathrm{f}}$, the final inclination angle of the planet's orbit relative to the initial fiducial plane.} for a USP-forming three-planet system. Note that in the examples shown in Fig. \ref{fig:fig_theta}, the stellar spin period is $P_{\star} = 30$ days, corresponding to a case where the spin-orbit coupling is negligible ($\omega_{i\star}$ is small). We find that indeed, when the system crosses $\epsilon_{12} \simeq 1$ during orbital decay, large mutual inclinations can be excited between the innermost planet and its companion. Moreover, this excitation is larger when the ratio  $m_{1,0}/m_{2,0}$ (and thereby $L_1/L_2$) is smaller: we find that systems with $m_1 = M_{\oplus}$ and $m_2 = 17 M_{\oplus}$ achieved a maximum value of $\langle \theta_{12,\mathrm{f}}^2 \rangle^{1/2} \approx 16$ deg., compared to $\langle \theta_{12,\mathrm{f}}^2 \rangle^{1/2} \approx 11$ deg. for the case of $m_2 = 7 M_{\oplus}$. 

When there is more substantial spin-orbit coupling, the mutual inclination evolution is similar, except that the resonance occurs at higher values of $\epsilon_{12}$. To illustrate this, in Fig. \ref{fig:fig_Pstar} we show the same examples as Fig. \ref{fig:fig_theta}, except with the stellar spin period fixed at $P_{\star} = 1$ day, corresponding to strong spin-orbit coupling (large $\omega_{i\star}$). In this case, there is still a resonant excitation in the mutual inclination, except that it happens at much larger values of $\epsilon_{12}$.

This shift of the mutual inclination resonance to higher values of $\epsilon_{12}$ can be understood if we consider that resonance occurs when 
\begin{equation}
    \omega'_1 = \omega_{12} + \omega_{13} + \omega_{1\star} \simeq \omega'_2 =  \omega_{21} + \omega_{23} + \omega_{2\star}.
\end{equation}
When spin-orbit coupling is negligible, $\omega_{i\star} \simeq 0$, and the resonance criterion corresponds to $\epsilon_{12} = (1 - L_1/L_2)/(1 + L_1/L_2)$, which is close to unity for $L_1 \ll L_2$. However, as the spin-orbit coupling becomes stronger, the resonance condition becomes $\epsilon_{12} \simeq (1-L_1/L_2)/(1 + L_1/L_2) + \omega_{1\star}/(\omega_{12} + \omega_{21})$, so the critical $\epsilon_{12}$ increases as $\omega_{1\star}$ increases. 

Another way to look at the role of spin-orbit coupling is to consider what happens when $P_{\star}$ changes while fixing the other parameters. In Fig. \ref{fig:spin_plot}, we show the final value of $\theta_{12,\mathrm{f}}$ and $\theta_{1\star,\mathrm{f}}$ as a function of the value of $P_{\star}$ (fixed in time during the evolution) for a three-planet system under-going low-$e$ USP formation. The system has initial parameters chosen such that $\epsilon_{12,0} = 0.1$ and $\epsilon_{12,\mathrm{f}} = 1.1$. We find that when $P_{\star} \ge 20$ days, the system indeed undergoes a secular inclination resonance, reaching a final value of $\langle \theta_{12,\mathrm{f}}^2 \rangle^{1/2} \approx 16$ deg. However, as the stellar spin period decreases below $P_{\star} = 20$ days, there is a sudden transition and the final values of both $\langle \theta_{12,\mathrm{f}}^2 \rangle^{1/2}$ and $\langle \theta_{1\star,\mathrm{f}}^2 \rangle^{1/2}$ decrease to much lower values. In general, for systems with $\epsilon_{12,0} < 1$ and $\epsilon_{12,f} \sim 1$, the ``transition'' in Fig. \ref{fig:spin_plot} occurs when $P_{\star}$ reaches a value such that $\omega_{1\star,\mathrm{f} } \gtrsim \omega_{12,\mathrm{f} }$ at the end of orbital decay. In other words, if $\omega_{1\star,\mathrm{f}} \gtrsim \omega_{12,\mathrm{f}}$, then spin-orbit coupling will suppress any resonant mutual inclination excitation between the innermost planets.

In real systems, the stellar rotation period increases over time, thus the importance of the spin-orbit coupling depends on the timescale of the proto-USP orbital decay: if the orbital decay occurs well with-in a Gyr, then spin-orbit coupling can be important. Otherwise, the star would have already spun down by the time the final USP semi-major axis is reached, and the effect of spin-orbit coupling is small.

\section{Population Synthesis Model}
\label{sec:5}
We synthesize the results of sections \ref{sec:2} - \ref{sec:4} by performing a population synthesis calculation of USPs generated through the low-e migration mechanism. Given the inherent uncertainties in various population statistics (of both USPs and larger-period planets), the purpose of this study is not to accurately reproduce all the observed population of USPs. Instead, our goal is to illustrate the statistical trends that would be expected when USPs are generated by low-e migration.

The initial semi-major axis of the inner-most planet $a_1$ is drawn from a power-law distribution given by
\begin{equation}
    \frac{dN}{d \log{P_1}} \propto P^{1.5}
\end{equation}
in the interval [$P_{\mathrm{min}}, P_{\mathrm{max}}$], with $P_{\mathrm{max}} = 8$ days and $P_{\mathrm{min}} = $ 0.5, 1.0, 2.0 and 3.0 days in four separate experiments. The planet's mass $m_1$ is drawn from a log-uniform distribution between $0.5$ and $3.0 ~M_{\oplus}$. The inner planet's composition is assumed to be Earth-like, with a radius of $R_1 = (m_1/M_{\oplus})^{1/4} R_{\oplus} $ \citep{Zeng2016}; its tidal lag time $\Delta t_{L,1}$ is chosen to be 1000, 100 or 10s, corresponding to $Q_1 = 7, 70$ and 700 for $P_1 = 1$ day. The outer planet masses are drawn from a log-uniform distribution between 3 and $20 M_{\oplus}$. We assign these outer planets ($i \ge 2$) a rocky compositions with a H/He envelope comprising a few percent of its mass, with radii given by $R_i = R_{i,\mathrm{core}} + R_{i,\mathrm{env}} = R_{\oplus} [(m_i/M_{\oplus})^{1/4} + 1.5]$, and tidal lag times $\Delta t_{L,i} = 1$ or 10 sec, corresponding to $Q_i = 7\times 10^{3}, ~7\times 10^{4}$ for $P_i = 10$ days. The initial semi-major axis of the outer planets are given by the ratios $a_2/a_1$ and $a_3/a_2$, chosen independently on a log-uniform distribution between 1.41 and 3.0, corresponding to period ratios between 1.67 and 5.2. The initial eccentricities of all planets are equal to $\bar{e}$, which is chosen from a Rayleigh distribution with scale parameter $\sigma_e$ = 0.10. The initial complex inclinations $\mathcal{I}_j$ of each planet are chosen from a 2-D Gaussian distribution with mean $\mu = 0$ and variance $\sigma_{\theta} = \bar{e}/2$; the resulting ratio $e / \theta \simeq 2$ is consistent with equipartition of random velocities suggested by numerical simulations of accreting planetesimals \citep{Kokubo2002}. This choice of inclinations is equivalent to a Rayleigh distribution for $|\mathcal{I}_i|$ with scale parameter equal to $\bar{e}/2$ and with the complex argument uniformly distributed between 0 and $2\pi$. 
We include the effect of tidal decay due to stellar tides, as given by Eq. (\ref{eq:gamma_star}). The value of $Q'_{\star}$ is chosen to be $10^6$, $10^7$ or $10^8$. We adopt an initial stellar spin of $P_{\star, 0} = 8$ days, subject to the Skumanich law (Eq. \ref{eq:skumanich}) with $a_{\mathrm{MB}} = 3.2\times 10^{-14}$ yr$^{-1}$ such that the stellar spin period lengthens to $P_{\star} = 30$ days at $t = 5$ Gyr.  

\color{black} We account for the possibility of dynamically unstable systems. A system of $N$ planets on initially circular orbits is stable up to $\tau \equiv t/P_1$ orbits if the spacing satisfies $(a_{i+1} - a_{i}) \ge k_c R_H$, where $k_c$ is a parameter that depends on $N$ and $\log{\tau}$, and $R_H$ is the mutual Hill radius given by
\begin{equation}
    R_H = \left(\frac{a_i + a_{i+1}}{2} \right)\left(\frac{M_i + M_{i+1}}{3M_{\star}} \right)^{1/3}.
\end{equation}
For mildly eccentric systems, the same criterion as above can be applied, but instead of the semi-major axis difference $(a_{i+1} - a_{i})$ one should use the pericenter-apocenter distance $a_{i+1}(1-e_{i+1}) - a_{i}(1-e_i)$ \citep{Pu2015}. We adopt a value of $k_c \sim 7$ \citep{Smith2009}, applicable for $N = 3$ and $\tau \sim 10^8$ (the typical eccentricity damping timescale); when systems fail to meet this stability criterion, they are regarded as potentially dynamically unstable.

We find that systems can indeed become potentially dynamically unstable before forming USPs. 17.4\% of systems that formed USPs and 6.3\% of systems that did not form USPs became dynamically unstable at some point of their evolution; systems that form USPs are more likely to become unstable due to their larger initial eccentricities, so dynamical instability may be an impediment to USP formation, although the effect is minor. \color{black}

We evolve our systems for 10 Gyr. In some cases, the inner planet's semi-major axis can shrink to a value less than $R_{\star}$; when this occurs, we assume the planet is tidally disrupted and/or engulfed by the star, and we remove it from the system and halt the simulation.

\begin{figure}
\includegraphics[width=0.99\linewidth]{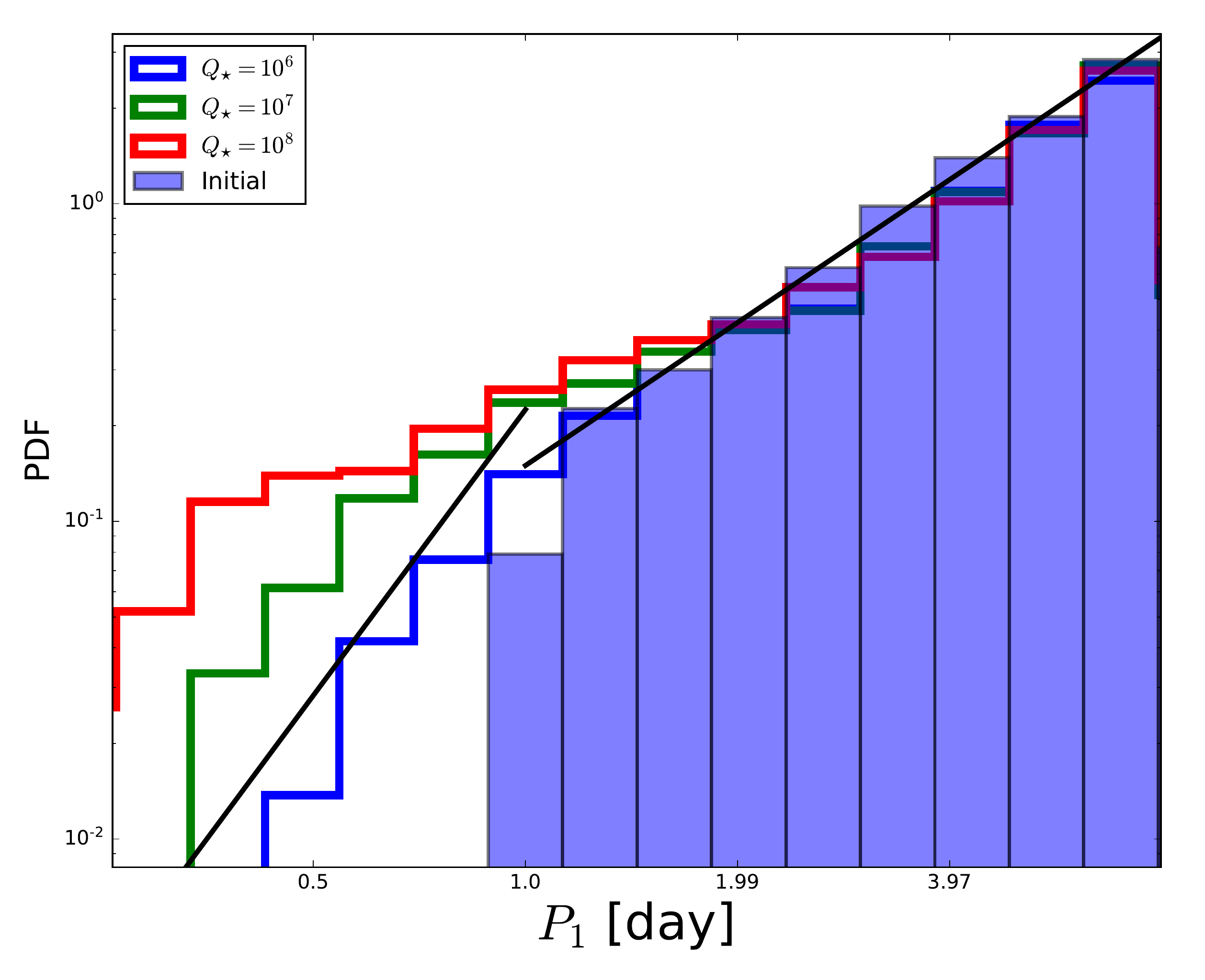}
\caption{Histogram of the initial and final period of the inner-most planet $P_1$ for systems with inner planet tidal lag time $\Delta t_{L,1} = 100$ s. The blue bars shows the initial distribution of $P_1$, while the blue, green and red lines show the final USP period distribution for values of $Q'_{\star} = 10^6, ~10^7$ and $10^8$ respectively. The two solid black lines are given by the power-law distribution $dN/d\log{P_1} \propto P_1^\alpha$, where $\alpha = 3.0$ for $P_1 \le 1 $ day and $\alpha = 1.5$ for $1 < P_1 < 8$ days; we also adopt the discontinuous ``bump'' at $P_1 = 1$ day corresponding to an excess of $50\%$ more planets just below $P = 1$ day as proposed by Lee \& Chiang (2017). The normalization of the black lines is chosen so that the total probability density integrates to unity. 
}
\label{fig:hist1}
\end{figure}

\begin{figure}
\includegraphics[width=0.99\linewidth]{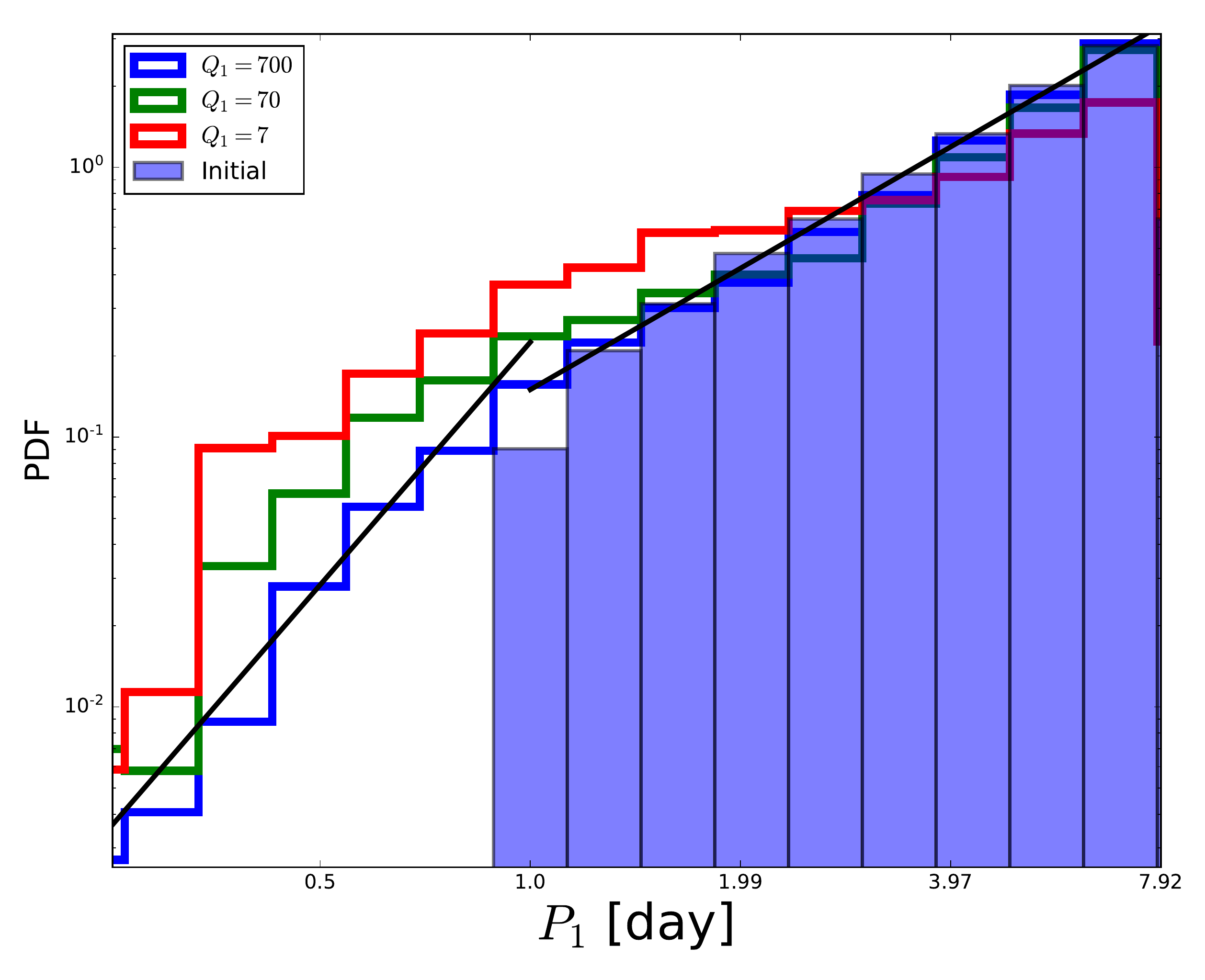}
\caption{Same as Fig. \ref{fig:hist1} except that we fix the value of $Q'_{\star} = 10^7$ and instead vary $\Delta t_{L,1} = 10, 100, 1000$ s, corresponding to tidal $Q_1 = 700, 70, $ and $7$ (at $P_1 = 1$ day), for the blue, green and red lines respectively.
}
\label{fig:hist2}
\end{figure}

\begin{figure}
\includegraphics[width=0.99\linewidth]{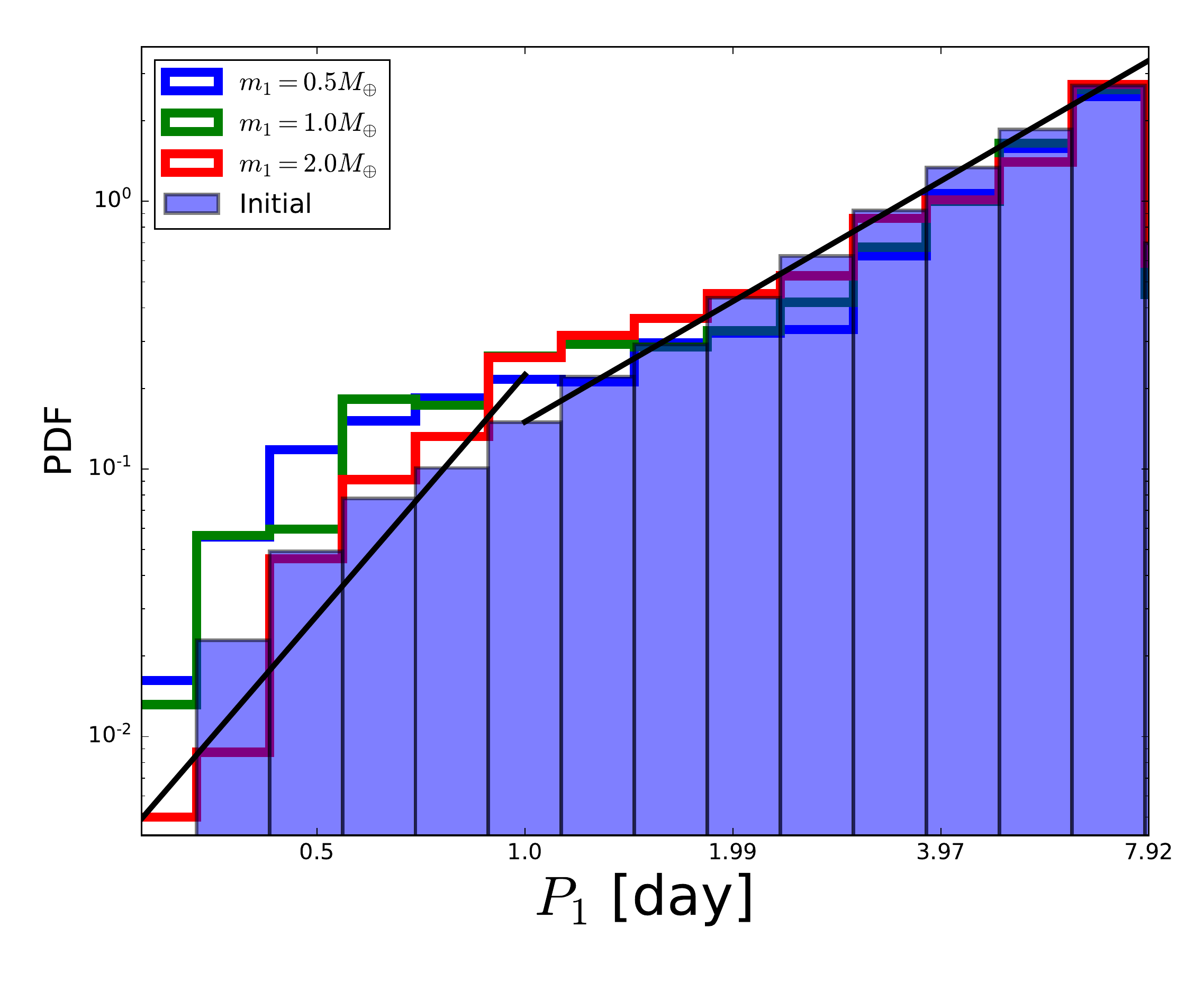}
\caption{Same as Fig. \ref{fig:hist1} except that we fix the value of $Q'_{\star} = 10^7$, $\Delta t_{L,1} = 100$ s and instead vary $m_1 = 0.5 \pm 0.25, 1.0 \pm 0.25$ and $2.0 \pm 0.25 ~M_{\oplus}$, for the blue, green and red lines respectively.
}
\label{fig:hist4}
\end{figure}

\begin{figure}
\includegraphics[width=0.99\linewidth]{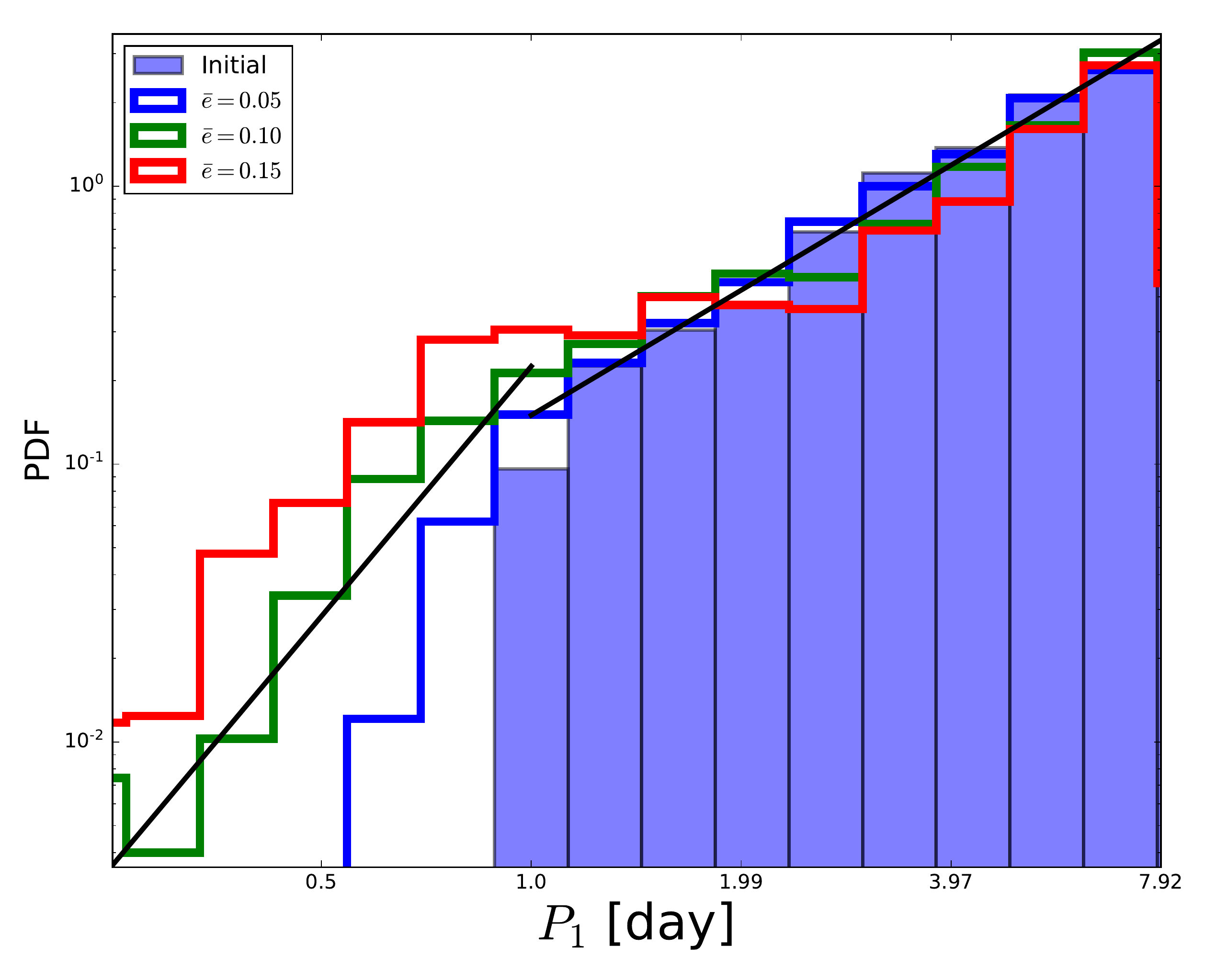}
\caption{Same as Fig. \ref{fig:hist1} except that we fix the value of $Q'_{\star} = 10^7$, $\Delta t_{L,1} = 100$ s and instead vary the initial eccentricity $\bar{e} = 0.05 \pm 0.025, 0.1 \pm 0.025,$ and $0.15 \pm 0.025$ for the blue, green and red lines respectively.
}
\label{fig:hist3}
\end{figure}

\begin{figure}
\includegraphics[width=0.99\linewidth]{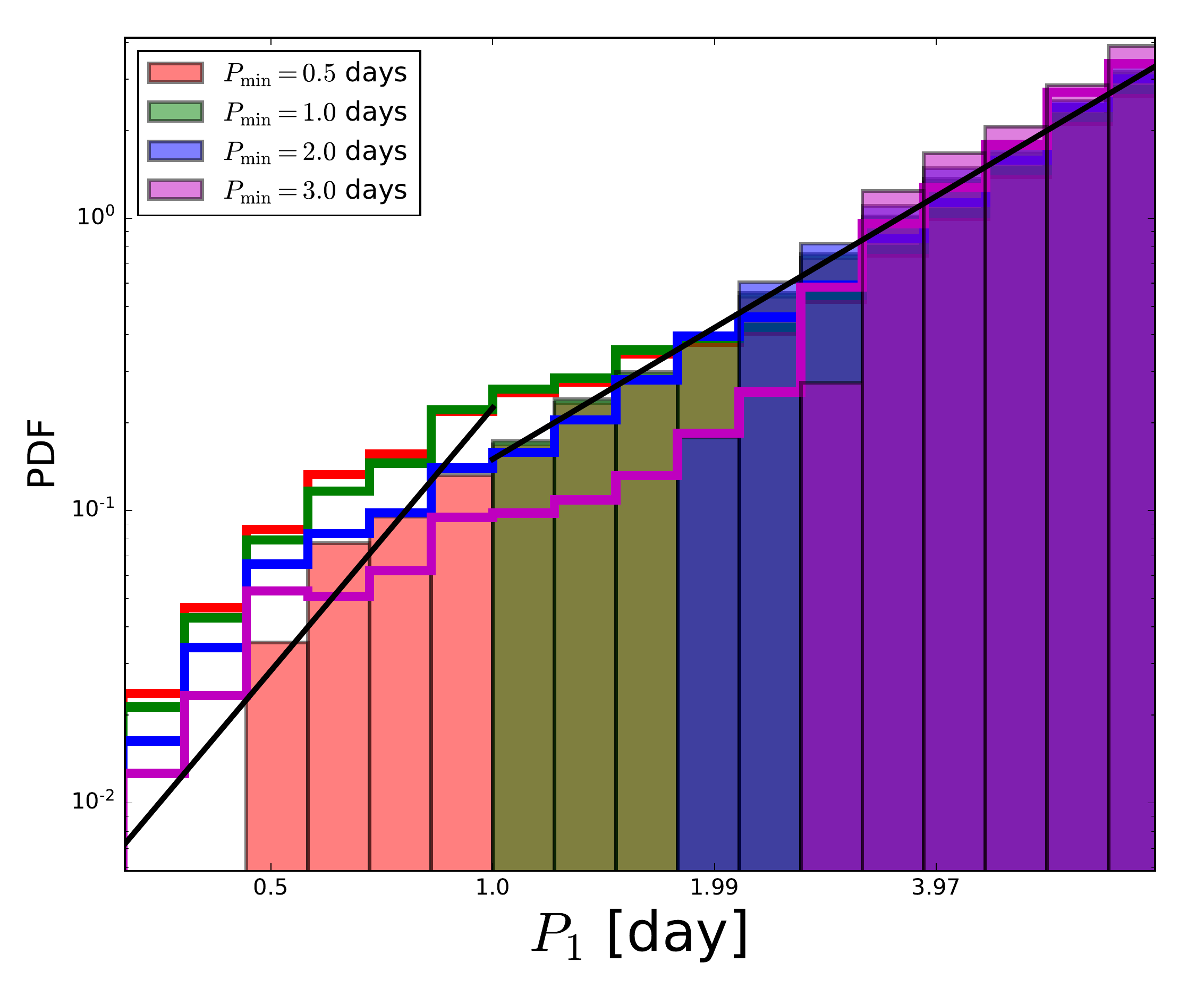}
\caption{Same as Fig. \ref{fig:hist1} except that we fix the value of $Q'_{\star} = 10^7$, $\Delta t_{L,1} = 100$ s and instead vary $P_{\mathrm{min}} = 0.5, 1.0, 2.0$ and 3.0 days (the minimum period of the initial planet population), for the red, green, blue and magenta colors respectively. The solid bars show the initial period distribution for the four values of $P_{\mathrm{min}}$ while the lines show the final distribution.
}
\label{fig:hist5}
\end{figure}

\begin{figure}
\includegraphics[width=0.99\linewidth]{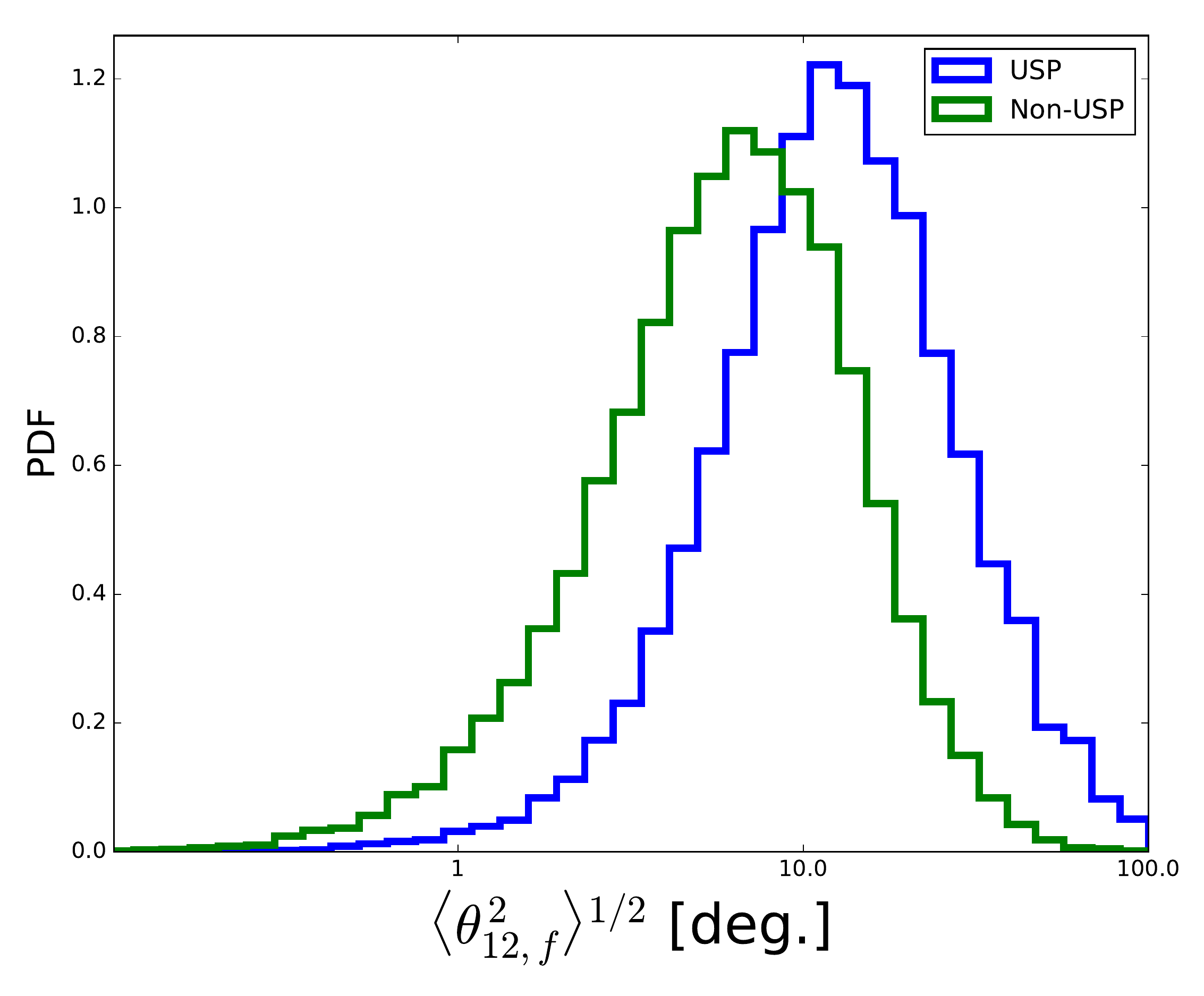}
\caption{PDF of the final RMS mutual inclination between the two inner planets after 10 Gyrs of integrations for all systems in our population synthesis. The blue line shows the mutual inclination for planets that became USPs, while the green line is for non-USPs. 
}
\label{fig:hist6}
\end{figure}

We found that our initial population of planet systems indeed formed USPs during its evolution, with statistical properties similar to the observed population. The USP population show substantial statistical differences with the longer period planets. We summarize their main properties below.
\begin{itemize}
\item {\bf Final period:} Our initial population of 3-planet systems is capable of producing USPs, with the final periods attained being as low as $P_1 \sim 5$ hrs. Most notably, we can reproduce both the sudden change in the slope of the period distribution at $P_1 \sim 1$ day and the mild excess of planets around $P_1 \sim 1$ day in our simulations (see Fig. \ref{fig:hist1}). This trend persists over a moderate range of of planet masses $m_1$, and inner planet tidal $Q_1$, but depends sensitively on the initial planet eccentricities $e_{2,0}, ~e_{3,0}$ (both equal to $\bar{e}$; see above) and the stellar tidal $Q'_{\star}$. Note that in our simulations, the final distribution for $P_1$ is not the same as the distribution for $P$ (of all the planets), since in some cases $P_2$ can also be in the range [1, 8] days, although this mixing does not affect our conclusions.
\item {\bf The value of $Q'_{\star}$:} The final period distribution of USPs has a strong dependence on the value of $Q'_{\star}$. We show a histogram of the initial and final periods for various choices of $Q'_{\star}$ in Fig. \ref{fig:hist1}. For systems with $Q'_{\star} = 10^6$, the period distribution of USPs is strongly carved by stellar tides over Gyr timescales, which results in fewer USP planets observed at smaller periods. On the other hand, systems with $Q'_{\star} = 10^8$ are not strongly affected by stellar tides, resulting in a much larger fraction of planets surviving at smaller periods. Our simulations suggest that for $\Delta t_{L,1} = 100$s, a value of $Q'_{\star}$ between $10^6$ and $10^7$ best matches the power-law period distribution of USPs given by \cite{Petigura2018} and \cite{Lee2017}, and $Q'_{\star} \gtrsim 10^8$ is incompatible with observations in our scenario.
\item {\bf The value of $Q_{1}$:} The final period distribution of USPs also depends moderately on the inner planet's tidal $Q_1$ (see Fig. \ref{fig:hist2}). We find that as expected, a larger value of $Q_1$ leads to fewer USPs: systems with $Q_1 = 70$ and 7 feature 2 and 3.5 times more USPs respectively than systems with $Q_1 = 700$.
\item {\bf Inner planet mass:} Less massive inner planets are more likely to become USPs. Over our sample, the inner planet's mass is smaller for USPs, with $\langle m_1 \rangle = 1.25 M_{\oplus}$ for USPs versus $1.5 M_{\oplus}$ for the entire population (see Fig. \ref{fig:hist4}). Systems with $0.5 M_{\oplus} < m_1 < 0.75 M_{\oplus}$ were 60\% more likely to form USPs than systems with $1.75 M_{\oplus} < m_1 < 2.25 M_{\oplus}$.
This is because USP formation is limited by the amount of angular momentum deficit (section \ref{sec:2.3}), and systems with less massive inner planets have an easier time reaching the required amount of AMD. 
\item {\bf Outer planet masses:} \color{black}Conversely, we find that USP production favors systems with more massive outer planets, although this is a weak effect. The average mass for the exterior planets across all samples is $11.5 M_{\oplus}$, and $12.1 M_{\oplus}$ for the subset that ended up producing USPs. \color{black}
\item {\bf Initial eccentricities:} We find that USP generation is strongly dependent on the initial eccentricities, with the fraction of systems producing USPs roughly doubling for every $0.05$ increase in $\bar{e}$: systems with $0.075 < \bar{e} < 0.125$ and $0.125 < \bar{e} < 0.175$ produce $1.7$ and $3.8$ times more USPs respectively than systems with $0.025 < \bar{e} < 0.075$. We show the dependence of the final period distribution for various initial eccentricities in Fig. \ref{fig:hist3}.
\item {\bf Initial inner period cutoff $P_{\mathrm{min}}$:} Our results can be used to constrain the minimum period $P_{\mathrm{min}}$ for the initial planet population. We show the dependence of the USP period distribution on $P_{\mathrm{min}}$ in Fig. \ref{fig:hist5}. The systems with $P_{\mathrm{min}} = 0.5$ day show little difference compared to those with $P_{\mathrm{min}} = 1$ day, because virtually all planets with initial $P_1 \le 1$ day spiral into their host stars through a combination of planetary and stellar tidal dissipation. In other words, the \color{black}low-$e$ USP migration\color{black}  mechanism is not sensitive to planets with initial $P \lesssim 1$ day. On the other hand, the results of experiments with $P_{\mathrm{min}} = 2$ or $3$ days show a substantial deviation from the $P_{\mathrm{min}} = 1$ day case, and disagrees with the observed period distribution. Thus, planets must be formed in the $1 < P < 3$ days range to reproduce the currently observed period distribution of USPs, although we cannot rule out the possibility of planets forming in-situ at $P_1 \lesssim 1$ day.
\item {\bf Mutual inclinations:} Observationally, USPs show substantially larger mutual inclinations with their closest neighbors compared with typical Kepler multis \citep{Dai2018}. We find that our low-e formation mechanism for USPs naturally generates larger mutual inclinations between the inner planets. We show a histogram of the final RMS mutual inclinations between the inner planets $\theta_{12,\mathrm{f}}$ after 10 Gyr of low-e migration in Fig. \ref{fig:hist6}. The final value of  $\theta_{12,\mathrm{f}}$ for systems that produced USPs is $\langle \theta_{12,\mathrm{f}}^2 \rangle^{1/2} \approx 18$ deg., which is more than double the value of $\langle \theta_{12,\mathrm{f}}^2 \rangle^{1/2} \approx 8$ deg. for systems that did not end up producing USPs. \color{black} A condition for the $i$-th planet to transit its host star is that the orbital plane be inclined relative to the line of sight by less than $\arcsin{\left(\frac{R_i + R_{\star}}{a_i}\right)}$. Using this criterion, we find that 17.5\% of USPs had transiting companions, compared with 63.5\% of inner planets with $P_1 > 1$ day. This result is consistent with empirical studies, which found USPs to have a transiting companion fraction of 4 - 12 \%, compared to $43 - 59\%$ for small planets with $1 \le P \le 3$ days \citep{Weiss2018}.
\item {\bf Inner Pair Period Ratio:} We find that systems which resulted in USPs have substantially larger period ratios $P_2/P_1$: USP systems have a mean period ratio of $P_2/P_1 = 14$, whereas for non-USP systems the mean period ratio is only 3.5. Fig. \ref{fig:hist7} shows the PDF of the initial and final period ratios. We also find the period ratio $P_2/P_1$ increases as $P_1$ decreases: the mean period ratio is 4.0, 5.2 and 7.0 for $P_1 = 3$, 2 and 1 day respectively. This is consistent with the observation that USPs and their companions have period ratios $\ge 15$, while non-USPs have a broader period ratio range between $1.4 \lesssim P_2/P_1 \lesssim 5$ \citep{Petrovich2018}.
\end{itemize}

\begin{figure}
\includegraphics[width=0.99\linewidth]{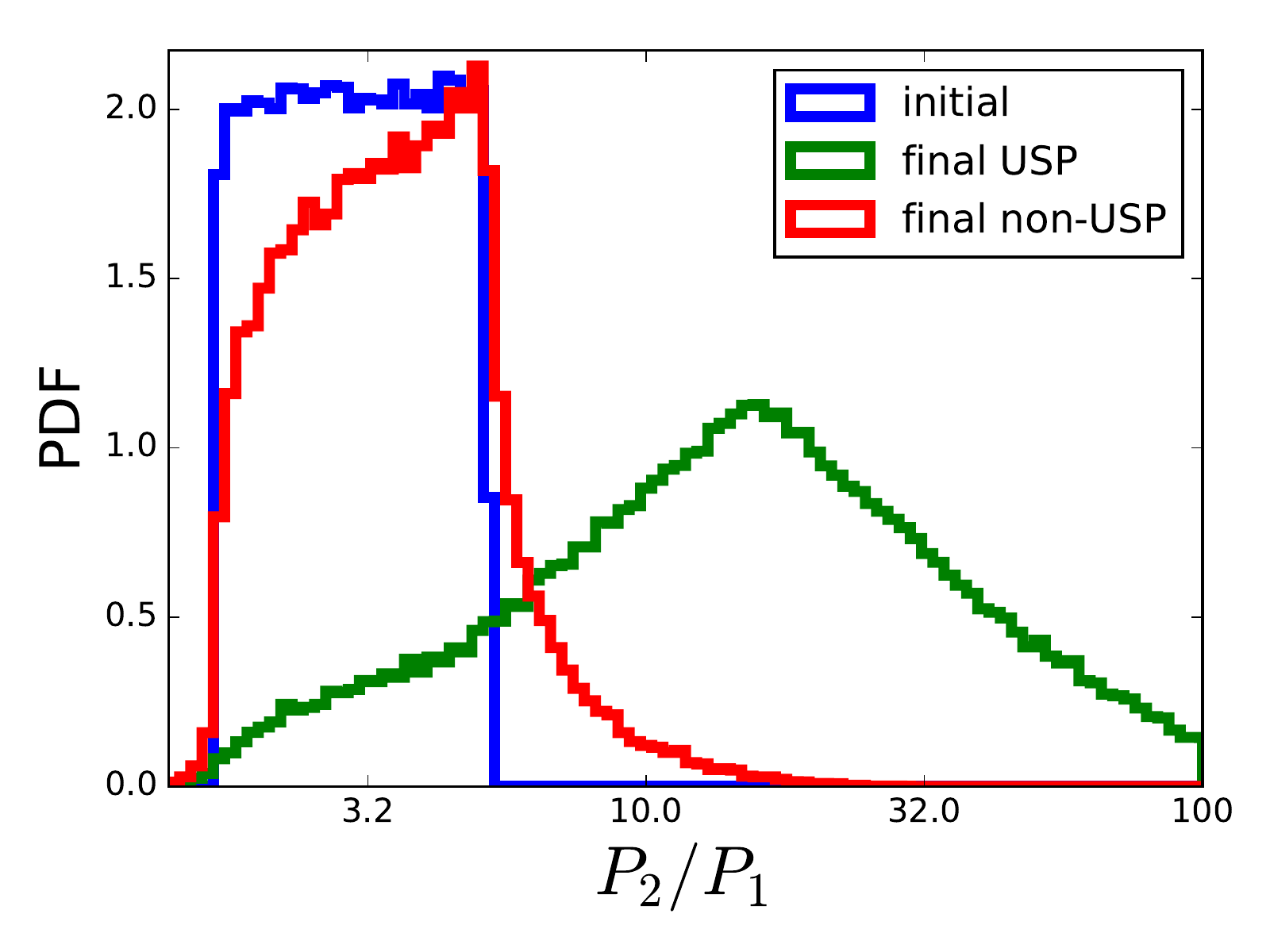}
\caption{PDF of the final initial and final period ratio of the inner planets $P_2/P_1$ distribution for systems in our population synthesis. The blue curve is the initial period ratio, while the green and blue curves are the final period ratios for systems that resulted in USPs and no USPs respectively.
}
\label{fig:hist7}
\end{figure}

\section{Discussion}
\label{sec:6}
We have studied the formation of USPs through low-$e$ tidal dissipation driven by secular forcings from exterior (super-Earth/mini-Neptune) companions of proto-USPs. In this section we evaluate this proposed formation mechanism in light of the observations of USPs and their population statistics. We then discuss some specific USP sources, potential uncertainties and future extensions to this work. 

\subsection{\color{black}Low-$e$ USP migration and observations \color{black}}
As discussed in section \ref{sec:1}, USPs have a number of distinct properties compared to the bulk of longer-period Super-Earth systems \citep[see][]{Winn2018}. Our study shows that our low-$e$ migration scenario produce USPs with the observed properties under a variety of initial conditions (see section \ref{sec:5}). USPs are preferentially formed from smaller terrestrial planets with more eccentric external companions. A fiducial set of systems, with inner planet masses $M_{\oplus} < m_1 < 3 M_{\oplus}$, exterior planet masses $3 M_{\oplus} < m < 20 M_{\oplus}$, inner planet $\Delta t_{L,1} = 10 $s (corresponding to $Q_1 = 700$ at $P_1 = 1$ day) and $Q'_{\star} = 10^7$ ended up producing a posterior USP period distribution that qualitatively matched the observed one, without any fine-tuning. Note that this combination of parameters is not the only one that can produce the observed $P_1$ distribution; there is a hyper-surface of possible initial system parameters that can fit the observations. For example, a set of systems with $Q'_{\star} = 3\times 10^6$ and $\Delta t_{L,1} = 100$ s would fit the observations similarly well. Nevertheless, the fact that our set-up was able to reproduce observations without tuning of parameters, and that similar looking distributions can be obtained when varying the parameters $Q_1$, $Q'_{\star}$, $m_1$ and $\bar{e}$ (see Figs. \ref{fig:hist1} - \ref{fig:hist5}) over factors of a few lends us confidence in the robustness of this mechanism.

Even more encouragingly, this formation mechanism naturally produces higher mutual inclinations between USPs and their closest companions, a trend which has been observed by empirical studies. \cite{Petrovich2018} found that a mutual inclination of $\langle \theta_{12,\mathrm{f}}^2 \rangle^{1/2} \gtrsim 20$ deg is needed to account for the observed dearth of transiting companions to USPs. In our population synthesis model, we found that systems that produced USPs featured a value of $\langle \theta_{12,\mathrm{f}}^2 \rangle^{1/2} \approx 18$ deg, which more than doubles the amount for systems that did not end up with USPs and is comparable to the value required by empirical studies. \color{black} USPs formed in our mechanism have a transiting companion fraction of 18\% compared with 64\% for planets with $1 \le P_1 \le 3$ days, close to empirical values of 4-12\% and 43-59\% respectively \citep{Weiss2018}. We also reproduce the observation that the period ratio between USPs and its companions tends to be large ($P_2/P_1 \gtrsim 15$) and increases with decreasing $P_1$ \cite[see also][]{Steffen2013}. \color{black}

The feasibility of the low-$e$ formation mechanism for USPs hinges mostly on the magnitude of the initial eccentricities of multi-planet systems: for systems with initial proto-USP periods between 1 - 3 days, our population model suggests that an initial eccentricity of $\bar{e} \gtrsim 0.1$ is required. In contrast, present-day Kepler multis have typical eccentricities of $\sigma_e = 0.05 - 0.08$ \citep{vanEylen2015,Xie2016,VanEylen2018}, although with the caveat that the presently observed planet eccentricities may have suffered damping over Gyrs by tidal dissipation, and their initial values may be larger. 


It is interesting to compare our mechanism with the secular chaos ``high-eccentricity'' mechanism proposed by \cite{Petrovich2018}. These two mechanisms require different initial conditions and produce USPs with distinct final configurations. In the \cite{Petrovich2018} scenario, proto-USPs with $a_{1,0}$ between 0.05 - 0.1 au attain large eccentricities ($1 - e_1 \ll 1)$ through to secular interactions with exterior planets; as the proto-USP pericenter reaches $\sim 2 R_{\odot}$, the planet is tidally captured and eventually circularized, becoming an USP. In contrast, our low-$e$ migration mechanism requires a proto-USP with $a_{1,0}$ between $0.02 - 0.04$ au, driven to mild eccentricities ($e_1 \sim 0.05 - 0.2$) through secular interactions with exterior planets, followed by tidal decay. USPs formed via secular chaos have a smaller ratio $a_{1,\mathrm{f}}/a_{1,0}$, and therefore a larger amount of AMD is needed (see Eq. \ref{eq:a1_min3pl}): one typically requires $N \ge 3$ exterior planets with  $m_i \gtrsim 10 M_{\oplus}$, $e_i \gtrsim 0.1$ and period ratios $P_{i+1}/P_{i} \gtrsim 3.0$, and planet companions with $a_2 \lesssim 0.2$ au are ruled out due to dynamical instability. In contrast, our low-$e$ migration mechanism requires less AMD to succeed: $N \ge 2$ exterior planets with $m_i \gtrsim 3 M_{\oplus}$ can satisfy the AMD constraint, and there is no need for large values of $P_{i+1}/P_i$. Observations \citep[e.g][]{Steffen2015} suggest that Kepler multis have typical period ratios $1.4 \le P_{i+1}/P_{i} \le 3.0$, and do not support the existence of large numbers of sparsely-spaced (i.e. $P_{i+1}/P_i \gtrsim 3.0$) multi-planet systems. In addition, while \cite{Petrovich2018} did not attempt to reproduce the final period distribution of USPs formed in their scenario, the low-$e$ migration mechanism can robustly reproduce the observed period distribution over a range of initial parameters (see section \ref{sec:5}).

Another difference between the high-$e$ and low-$e$ migration is the final distribution of the USP inclination ($\theta_1$). For a system under-going secular chaos, whenever $e_1$ grows to a very large value so too will the value of $\theta_1$ due to the equipartition principle \cite[e.g.][]{Lithwick2014}. In the absence of strong spin-orbit coupling, \cite{Petrovich2018} found that USPs can often reach very large values of inclination, with potentially a third of systems attaining $\theta_1 \ge 30$ deg. In contrast, the low-$e$ migration scenario produces USPs with more mild inclinations ($\theta_1 \sim 18$ deg), although this value is still enhanced relative to non-USPs.

\subsection{Specific sources}
\label{sec:specific_sources}
We comment below on the feasibility of \color{black} USP low-$e$  migration \color{black} for two well-studied USP systems.
\begin{itemize}
{\item {\bf Kepler 10} \citep[with $M_\star = 0.91 M_{\odot}$, $R_\star = 1.065 R_{\odot}$,][]{Batalha2011b, Fressin2011} is a system with 2 transiting planets: Kepler-10b is an USP with $m_1 = 3.72 M_{\oplus}$, $R_1 = 1.47R_{\oplus}$ and $a_1 = 0.0168$ au, while Kepler-10c is a sub-Neptune with $m_2 = 7.37 M_{\oplus}$, $R_2 = 2.35 R_{\oplus}$ and $a_2 = 0.24$ \citep{Rajpaul2017}. Kepler-10b has an inclination $\theta_{12} = 5.2$ deg relative to the orbital plane of Kepler-10c. TTV analysis suggests the existence of a third, non-transiting planet (Kepler-10d) with $a_3 = 0.366$ au and $m_3 = 7 M_{\oplus}$ \citep{Weiss2016}. \cite{Petrovich2018} found that in order for Kepler-10b to migrate to its current orbit from $a_{1,0} = 0.1$ au through high-$e$ migration, one requires three additional Neptune-mass planets with periods of about 122, 480 and 2100 days.

We found that low-$e$ migration can naturally reproduce Kepler-10b's current orbit, if one hypothesizes an additional fourth planet (Kepler-10e) located between Kepler-10b and Kepler-10c. For example, an initial configuration with $\Delta t_{L,1} = 100$ s, $a_{1,0} = 0.035$ au, $e_{2,0} = e_{3,0} = e_{4,0} = 0.2$, $m_4 = 7 M_{\oplus}$ and $0.074 \le a_{4,0} \le 0.121$ au can reproduce the final orbit of Kepler-10b. This configuration makes Kepler-10b and Kepler-10e ``tightly-coupled'', such that their orbits are aligned with each other and misaligned relative to Kepler-10c and Kepler-10d ($\theta_{42} \simeq \theta_{43} \sim 5$ deg). In this scenario, the fact that the hypothetical Kepler-10e would fail to transit is compatible with observations.

}
{\item {\bf Kepler-290}  \citep[with $M_\star = 0.8 M_{\odot}$, $R_\star = 0.7 R_{\odot}$,][]{Rowe2014} has a transiting USP accompanied by two outer planets. The USP (KOI 1360.03) with $R_1 = 0.97 R_{\oplus}$ and $a_1 = 0.151$ was discovered by \cite{Sanchis-Ojeda2014} and not formally vetted by {\it Kepler}; the outer planets have $a_2 = 0.11$ au, $a_3 = 0.205$ au, $R_2 = 2.7 R_{\oplus}$ and $R_3 = 2.3 R_{\oplus}$. Given the location and mass of the outer planets, this system can naturally produce USPs ``out-of-the-box'' through low-$e$ migration: assuming masses $m_1 = M_{\oplus}$, $m_2 = 9M_{\oplus}$, $m_3 = 7 M_{\oplus}$ and $\bar{e} = e_{2,0} = e_{3,0} = 0.2$, the final system can be reproduced as long as $a_{1,0} \lesssim 0.038$ au and $Q'_{\star} \lesssim 6\times10^6$. Decreasing the initial eccentricity to $\bar{e} = 0.15$ would instead require $a_{1,0} \lesssim 0.032$ au. The value of $a_{1,\mathrm{f}}$ depends moderately on $a_{1,0}$ but is highly sensitive to $\bar{e}$ and $Q'_\star$.
}
\end{itemize}

\subsection{Are USPs photo-evaporated cores of mini-Neptunes?}
The observed population of USPs have radii that are mostly within the range $1.0R_{\oplus} < R < 1.4R_{\oplus}$, and there is a dearth of planets with intermediate radius $2R_{\oplus} < R < 4R_{\oplus}$ with sub-day periods, despite such planets being ubiquitous amongst Kepler's longer-period planet population. One common explanation for this observation is the scenario that USPs were initially mini-Neptunes that have had their envelopes stripped due to photo-evaporation \cite[e.g.][]{Winn2017}. This picture may be incompatible with our model, and an alternative explanation might be preferred. Our results show that because low-$e$ USP formation is generally AMD-limited (section \ref{sec:3.3}), more massive planets are severely disfavored from becoming USPs. This would preclude higher-mass super-Earths or mini-Neptunes from becoming USPs. As a result, USP formation is limited to smaller mass planets ($m_1 \lesssim M_{\oplus}$), which would have a hard time maintaining their atmospheres against various escape mechanisms.

Another factor that can potentially explain the lack of larger-radius USPs is the dichotomy in tidal $Q_1$ between rocky planets and those with more extended gaseous envelopes. Our results show a reduction in USP formation efficiency by a factor of $\sim 2$, when $Q_1$ is increased by a factor of 10. In the Solar System, values of tidal $Q_1$ are in the range of $10 - 500$ for terrestrial planets and satellites, but planets with substantial gaseous envelopes (such as Jupiter, Saturn, Uranus and Neptune) have values of $Q_1$ that are hundreds of times larger \citep{Goldreich1996, Lainey2016}. If this trend can be extrapolated to exoplanetary systems, then this dichotomy in tidal $Q_1$ between planets with and without gaseous envelopes can also explain the lack of USPs with $R_1 \gtrsim 2 R_{\oplus}$.

\subsection{Uncertainties and Future Work}
In carrying out this work, we made several simplifications, which may cause additional uncertainties; we discuss them below.
\begin{itemize}
    \item {\bf Effects of Mean Motion Resonance:}
One source of uncertainty is the role of mean-motion resonance (MMR) in modulating the secular interactions between planets. In our population synthesis model, we considered planet systems with semi-major axes ratios $1.41 \le a_2 / a_1 \le 3$. In many cases, as the inner planet migrates in-wards the system may encounter MMRs \citep[see also][]{Hansen2015a}. A careful study of the effect of MMRs on the secular interactions of multi-planet systems is beyond the scope of this work. MMRs can excite the eccentricities of the planets, independent of secular interactions. One example is Kepler-80, a system containing an USP accompanied by 5 external planets. \cite{MacDonald2016} found that the outer 4 planets of Kepler-80 are interlocked in 4 sets of three-body mean-motion resonances, each with a libration of around a few degrees. The resulting librations may provide the entire system with an additional source of AMD that ameliorates the effect of tidal dissipation. Another possibility is that MMRs can result in `resonant repulsion', which would cause the semi-major axes of the two planets in resonance to suddenly diverge outside of the MMR \citep{Batygin2013, Lithwick2012b}. MMRs could bring about unexpected and interesting interactions in proto-USP systems and deserves to be the subject of further study.

\item{{\bf Secular Chaos and Dynamical Instability:}
In this work, we adopted a linear theory in the planet eccentricities and inclinations (by assuming $e_i, ~\theta_i \ll 1$). In this linear regime, the eccentricity and mutual inclination evolution of the planet orbits are  decoupled. In reality, planet systems that produce USPs will often have inner planets with moderately large values of $e_1 \gtrsim 0.3$. Such values would make higher-order terms in $e$ and $\theta$ important, and our linear theory would break down. A non-linear coupling between planet eccentricities and inclinations can bring about secular chaos \citep{Lithwick2014}, which can enhance the inner planet eccentricities even further as AMD diffuses throughout the system. 

Another issue is that as planet eccentricities increase, their orbits may become dynamically unstable leading to orbit crossings. \color{black}In our population study, we found that a small proportion ($\sim 17\%$, Sec. \ref{sec:5}) of systems that became USPs may become dynamically unstable. \color{black} For these inner systems, the dominant final outcome of dynamical instability is physical collision between the two unstable planets.  Once two planet have crossing orbits, for large eccentricities and inclinations (i.e. $e_1, \theta_{12} \gg [(m_1+m_2)/3M_{\star}]^{1/3}$) the timescale to the first physical collision is given by \citep[e.g.][]{Ida1989}:
\begin{equation}
    T_{\mathrm{coll}} \sim P_1 \left(\frac{R_1+R_2}{a_1} \right)^{-2} = 700 \left(\frac{a_1}{0.03 \mathrm{au}} \right)^{7/2} \left(\frac{R_1+R_2}{2 R_{\oplus}}\right)^{-2} ~\mathrm{yr}.
\end{equation}

Since the collisional timescale is much shorter than the eccentricity damping and orbital decay timescale, once two planets cross orbits, they will quickly undergo a physical collision, which can potentially inhibit USP formation. The extent to which these dynamical instabilities occur requires investigations using numerical N-body simulations and is outside the scope of this work.
}

\item{ {\bf Effect of Additional Planets:}
In this work we have limited our attention to USP formation in systems with 2 or 3 planets. What happens when additional planets are present? Our framework for 3-planet proto-USP systems can be easily generalized to systems with more than 3 planets. In general, the generation of USPs is constrained by the dual criteria that the system must have sufficient AMD (Eq. \ref{eq:ecrit_1}), and the forced eccentricity on the inner planet must be sufficiently large (Eq. \ref{eq:ecrit_2}). In section \ref{sec:3}, we found that for planet systems with $\bar{e} \gtrsim 0.1$, the AMD criterion is usually more stringent. The presence of additional exterior planets only help to overcome this constraint and bolster the chances of USP generation, since having more outer planets will increase the total reservoir of AMD to maintain the tidal decay of the inner planet. Moreover, the presence of additional planets (and thereby eigenmodes) increases the likelihood of hitting one of eccentricity secular resonances that can speed up the tidal orbital decay timescale. As a result, we expect USP formation in systems with $N \ge 4$ planets to be similar to systems with $N = 3$ planets, albeit at an enhanced rate.}

\end{itemize}

\section{Summary and Conclusion}
\label{sec:7}
In this paper we have studied a ``low-eccentricity'' migration scenario for the formation of USPs. In this scenario, a low-mass ($m_1 \sim M_{\oplus}$) inner planet with initial period of a few days is accompanied by several external planets in configurations typical of Kepler multi-planetary systems; the companion planets excite and maintain the eccentricity of the innermost planet, which then experiences orbital decay due to tidal dissipation and eventually becomes a USP. Tidal dissipation in the host star further enhances this orbital decay when the inner planet reaches a sufficiently small period. We find that this low-$e$ mechanism naturally produce USPs from the large population of Kepler multis, and can explain most of the observed population properties of USPs. The key findings of this paper are:

$\bullet$ We study analytically the condition for orbital decay of the inner planet induced by secular forcing from the outer planetary companions for systems with $N = 2$ (section \ref{sec:2.2}) or $N = 3$ (section \ref{sec:3}) planets. USP formation is governed by two criteria (section \ref{sec:2.3}): (i) the total system angular momentum deficit (AMD) must be sufficiently large, and (ii) the forced eccentricity on the inner planet must be sufficiently large so that decay occurs within the lifetime of the system. We find that it is difficult for 2-planet systems to simultaneously satisfy both criteria due to the suppression of forced eccentricity on the inner planet by short-range forces. On the other hand, 3-planet systems have a much easier time forming USPs (section \ref{sec:3}), as the presence of the 3rd planet introduces secular resonances that can boost the inner planet eccentricity, in addition to enhancing the AMD reservoir.

$\bullet$ Although the basic equations (based on secular Laplace-Lagrange theory) for eccentricity excitation and orbital decay in multi-planet systems are standard, in practice they are computationally difficult to evolve for long periods of time due to the ``stiffness'' of the equations: whereas orbital decay occurs on Gyr timescales, secular interactions proceed on timescales as short as $\sim 100$ yr. To resolve this, we develop an approximate method based on the evolution of eigenmodes (section \ref{sec:2.1}). We find that eigenmode crossing during orbital decay can lead to secular resonances, which can excite large eccentricities in the inner planet.

$\bullet$ We extend our analysis to the mutual inclination evolution in section \ref{sec:4}. We find that secular inclination resonances can also excite mutual inclinations between the innermost planet and its companions as it undergoes tidal decay. Moreover, the range of parameters for which the secular inclination resonance and secular eccentricity resonance occur usually coincides with one another, which results in large mutual inclinations being generated whenever a USP is formed.

$\bullet$ Using our approximate ``eigenmode'' method, we carry out a large population synthesis study to examine the statistical properties of USPs formed in the low-$e$ migration scenario (section \ref{sec:5}). We find that USPs can be robustly produced from typical Kepler multis under a range of initial conditions. This formation mechanism favors smaller inner planets, and requires the initial eccentricities of the companion planets to be $\bar{e} \gtrsim 0.1$. We find that the final USP period distribution depends on the values of planet tidal $Q_1$ and stellar tidal $Q'_{\star}$; in particular, a configuration with proto-USP mass $M_{\oplus} < m_1 < 3 M_{\oplus}$ and tidal lag time $\Delta t_{L,1} \sim 10 - 1000$s, outer planet masses $3 M_{\oplus} < m_i < 20 M_{\oplus}$ ($i \ge 2)$ and $Q'_{\star} \sim 10^6 - 10^7$ produces USPs with a final period distribution that matches closely with the observed one.

$\bullet$ Confronting with observations, we find that our low-$e$ migration mechanism can reproduce the empirical population properties of USPs. The final period distribution of USPs matches with the empirical distribution, and the radius distribution of USPs are biased towards small, Earth-like planets, in agreement with observations. Moreover, we find that in our low-$e$ formation mechanism, systems with USPs have more than twice as large mutual inclinations between the innermost planets as do systems without USPs, in agreement with other empirical studies. \color{black} Our mechanism also reproduces the empirical fraction of USPs with transiting companions, as well as the period ratio distribution of such USPs, without fine tuning of initial parameters.\color{black}

Overall, we conclude that the low-$e$ migration mechanism can more robustly produce the observed USPs than some of the other proposed mechanism (see section \ref{sec:6}). For some systems (e.g. Kepler-10), our scenario makes specific predictions for the existence of unseen planets which can be tested by future observations.

\section*{Acknowledgements}
\color{black}We thank Cristobal Petrovich, Fei Dai and Lauren Weiss for helpful discussions, as well as suggestions made by an anonymous referee. This work is supported in part by NSF grant AST1715246 and NASA grants NNX14AP31G and NNX14AG94G. BP is supported by a NASA Earth and Space Sciences Fellowship.
\color{black}

\bibliography{msNotes}
\bibliographystyle{mnras}

\end{document}